%% file: report.tex
\documentclass[
	a4paper,					
	10pt,							
	twoside,					
	openright,				
	notitlepage,			
	parskip=half,			
]{scrreprt}					

\KOMAoptions{cleardoublepage=plain}			

\usepackage{draftwatermark}
\SetWatermarkText{}

\usepackage[standard-baselineskips]{cmbright}

\usepackage[english]{babel}						
\usepackage[utf8]{inputenc} 
\usepackage[T1]{fontenc}						
\usepackage{textcomp}							
\usepackage{ae}									
\usepackage{fancyhdr}							
\usepackage{etoolbox}							
\usepackage{graphicx}                      		
\usepackage{float}								
\usepackage{caption}							
\usepackage{booktabs}							
\usepackage{tocvsec2}							
\usepackage{pdfpages}							

\usepackage{amsmath}                    	   	
\usepackage{amsthm}                       	 	
\usepackage{amsfonts}                      		
\usepackage{amssymb}							
\usepackage{exscale}							

\usepackage[absolute]{textpos}
\RequirePackage{color}                          
\definecolor{linkblue}{rgb}{0,0,0.8}            
\definecolor{darkblue}{rgb}{0,0.08,0.45}        
\definecolor{bfhgrey}{rgb}{0.41,0.49,0.57}      
\definecolor{linkcolor}{rgb}{0,0,0}        		

\usepackage[
	pdftex,ngerman,bookmarks,plainpages=false,pdfpagelabels,
	backref = {false},										
	colorlinks = {true},                  
	hypertexnames = {true},               
	bookmarksopen = {true},               
	bookmarksopenlevel = {0},             
	pdftitle = {DIGNITE},	   			  
	pdfauthor = {endet2; vanap1},		  
	pdfsubject = {Report for Bachelor Thesis},  
	linkcolor = {linkcolor},              
	citecolor = {linkcolor},              
	urlcolor = {linkcolor},               
]{hyperref}

\usepackage{geometry}
\usepackage{makeidx}                         		
\makeindex                                    	

\usepackage[nonumberlist]{glossaries}
\makeglossaries
\include{datenbanken/glossar}

\usepackage{listings} 

\definecolor{dkgreen}{rgb}{0,0.6,0}
\definecolor{gray}{rgb}{0.5,0.5,0.5}
\definecolor{mauve}{rgb}{0.58,0,0.82}

\begin{document}                              	
\settocdepth{subsection}														
\pagenumbering{roman}														

\input{vorspann/titel}					
\input{vorspann/version}				

\makeatletter
\patchcmd{\@fancyhead}{\rlap}{\color{bfhgrey}\rlap}{}{}		
\patchcmd{\@fancyfoot}{\rlap}{\color{bfhgrey}\rlap}{}{}		
\makeatother

\fancyhf{}																		
\fancypagestyle{plain}{												
	\fancyfoot[OR,EL]{\footnotesize \thepage} 	
	\fancyfoot[OL,ER]{\footnotesize \titel, Version \versionnumber, \versiondate}	
}

\renewcommand{\chaptermark}[1]{\markboth{\thechapter.  #1}{}}
\renewcommand{\headrulewidth}{0pt}				
\renewcommand{\footrulewidth}{0pt} 				

\pagestyle{plain}
\include{vorspann/titelseite_mit_bild}			
\include{vorspann/versionen}
\cleardoubleemptypage
\setcounter{page}{1}
\cleardoublepage
\phantomsection 
\addcontentsline{toc}{chapter}{Management Summary}
\include{vorspann/management_summary}
\cleardoubleemptypage

\begingroup
\let\cleardoublepage\relax
\let\clearpage\relax

\tableofcontents
\cleardoublepage

\newpage
\cleardoublepage
\phantomsection 
\addcontentsline{toc}{chapter}{Index of images}
\listoffigures
\phantomsection 
\addcontentsline{toc}{chapter}{Index of tables}
\listoftables
\endgroup
\cleardoublepage
\phantomsection 
\addcontentsline{toc}{chapter}{Glossary}
\renewcommand{\glossaryname}{Glossary}
\printglossary
\cleardoublepage
\pagenumbering{arabic}

\include{kapitel/01-introduction}
\include{kapitel/02-selectedtraces}
\include{kapitel/03-api}
\include{kapitel/04-securityandperformance}

\include{kapitel/05-demoapp}
\include{kapitel/06-conclusion}

\cleardoublepage
\phantomsection 
\addcontentsline{toc}{chapter}{Bibliography}
\bibliographystyle{IEEEtranS}
\bibliography{datenbanken/bibliography}{}
\cleardoublepage
\phantomsection 
\addcontentsline{toc}{chapter}{Declaration of conformity}
\include{vorspann/selbstaendigkeitserklaerung}
\appendix
\settocdepth{section}
\include{anhang/A0-usecases-ssd}

\include{anhang/A3-scala}
\include{anhang/A1-meetings}

\include{anhang/A2-workinglog}
\include{anhang/A4-projecttwo}

\end{document}

%% file: datenbanken/glossar.tex
\newglossaryentry{RFC}{name={Request for Comments},description={a document containing specifications about technologies or protocols}}
\newglossaryentry{IP}{name={IP},description={Internet Protocol, communication protocol for relaying datagrams across networks}}
\newglossaryentry{DNS}{name={DNS},description={Domain Name System, hierarchical distributed naming system used to associate \gls{IP} address and domain name}}
\newglossaryentry{DNSSEC}{name={DNSSEC},description={Domain Name System, Security Extensions}}
\newglossaryentry{TSIG}{name={TSIG},description={Transaction Signature, used for DNS updates}}
\newglossaryentry{AXFR}{name={AXFR},description={DNS query type used to initiate a DNS zone transfer}}
\newglossaryentry{TLD}{name={Top Level Domain},description={The second-highest level of the DNS tree, like ".ch","com" or "de"}}
\newglossaryentry{IANA}{name={Internet Assigned Numbers Authority},description={presently a department of ICANN, a nonprofit private American corporation, which oversees global \gls{IP} address allocation, autonomous system number allocation, root zone management in the Domain Name System}}
\newglossaryentry{CA}{name={Certificate Authority},description={entity which issues digital certificates}}
\newglossaryentry{ISP}{name={ISP},description={Internet Service Provider}}
\newglossaryentry{Web crawler}{name={Web crawler},description={An Internet bot that systematically browses the World Wide Web, typically for the purpose of Web indexing}}
\newglossaryentry{SBL}{name={SBL},description={Spamhaus Block List}}
\newglossaryentry{XBL}{name={XBL},description={Exploits Block List}}
\newglossaryentry{DBL}{name={DBL},description={Domain Block List}}
\newglossaryentry{ROKSO}{name={ROKSO},description={Register of Known Spam Operations}}
\newglossaryentry{DNSBL}{name={DNSBL},description={DNS-based Blocklists}}
\newglossaryentry{RHSBL}{name={RHSBL},description={Right-Hand-Side Blackhole List},description={a \gls{DNSBL} containing domain names instead of IP}}
\newglossaryentry{URI}{name={URI},description={Uniform Resource Identifier}}
\newglossaryentry{URL}{name={URL},description={Uniform Resource Locator}}
\newglossaryentry{API}{name={API},description={Application programming interface}}
\newglossaryentry{Captcha}{name={Captcha},description={Completely Automated Public Turing-test to tell Computers and Humans Apart}}
\newglossaryentry{SEO}{name={SEO},description={Search-Engine Optimization, used to improve result ranking of a site}}
\newglossaryentry{ID}{name={ID},description={Identifier that uniquely identifies an Object or a Record}}
\newglossaryentry{SSL}{name={SSL},description={Secure Socket Layer, together with its successor \glossary{TLS} used to encrypt communication with websites}}
\newglossaryentry{TLS}{name={TLS},description={Transport Layer Security, successor of SSL, used to encrypt communication with websites}}
\newglossaryentry{SOLID}{name={SOLID},description={is a set of five guiding principles that help developers design objects that are easy to maintain and use}}
\newglossaryentry{HTML}{name={HTML},description={Hypertext Markup Language, describes webpages}}
\newglossaryentry{SSD}{name={SSD},description={System Sequence Diagram}}
\newglossaryentry{SD}{name={SD},description={Sequence Diagram}}
\newglossaryentry{XML}{name={XML},description={Extended Markup Language}}
\newglossaryentry{CI}{name={CI},description={Continuous Integration}}
\newglossaryentry{IDE}{name={IDE},description={Integrated Development Environment}}
\newglossaryentry{GUI}{name={GUI},description={Graphical User Interface}}
\newglossaryentry{MD2}{name={MD2},description={Message Digest 2}}
\newglossaryentry{JAP}{name={JAP},description={Java Anonymous Proxy, a project of TU Dresden}}
\newglossaryentry{HTTP}{name={HTTP},description={Hypertext Transfer Protocol}}
\newglossaryentry{HTTPS}{name={HTTPS},description={Hypertext Transfer Protocol, Secure}}
\newglossaryentry{CPU}{name={CPU},description={Central Processing Unit}}

%% file: vorspann/titel.tex
\providecommand{\titel}{Dignité}		
\providecommand{\untertitel}{DIGital Network Information \& Traces Extraction}		

%% file: vorspann/version.tex
\providecommand{\versionnumber}{1.0}			
\providecommand{\versiondate}{13.06.2014}		

%% file: vorspann/titelseite_mit_bild.tex
%
%

\begin{titlepage}

\setlength{\unitlength}{1mm}
\begin{textblock}{20}[0,0](28,12)
	\includegraphics[scale=1.0]{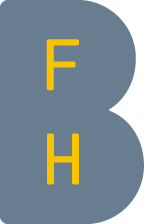}
\end{textblock}

\begin{textblock}{154}(28,48)
	\begin{picture}(150,2)
		\put(0,0){\color{bfhgrey}\rule{150mm}{2mm}}
	\end{picture}
\end{textblock}

\begin{textblock}{154}[0,0](28,50)
	\includegraphics[scale=0.83]{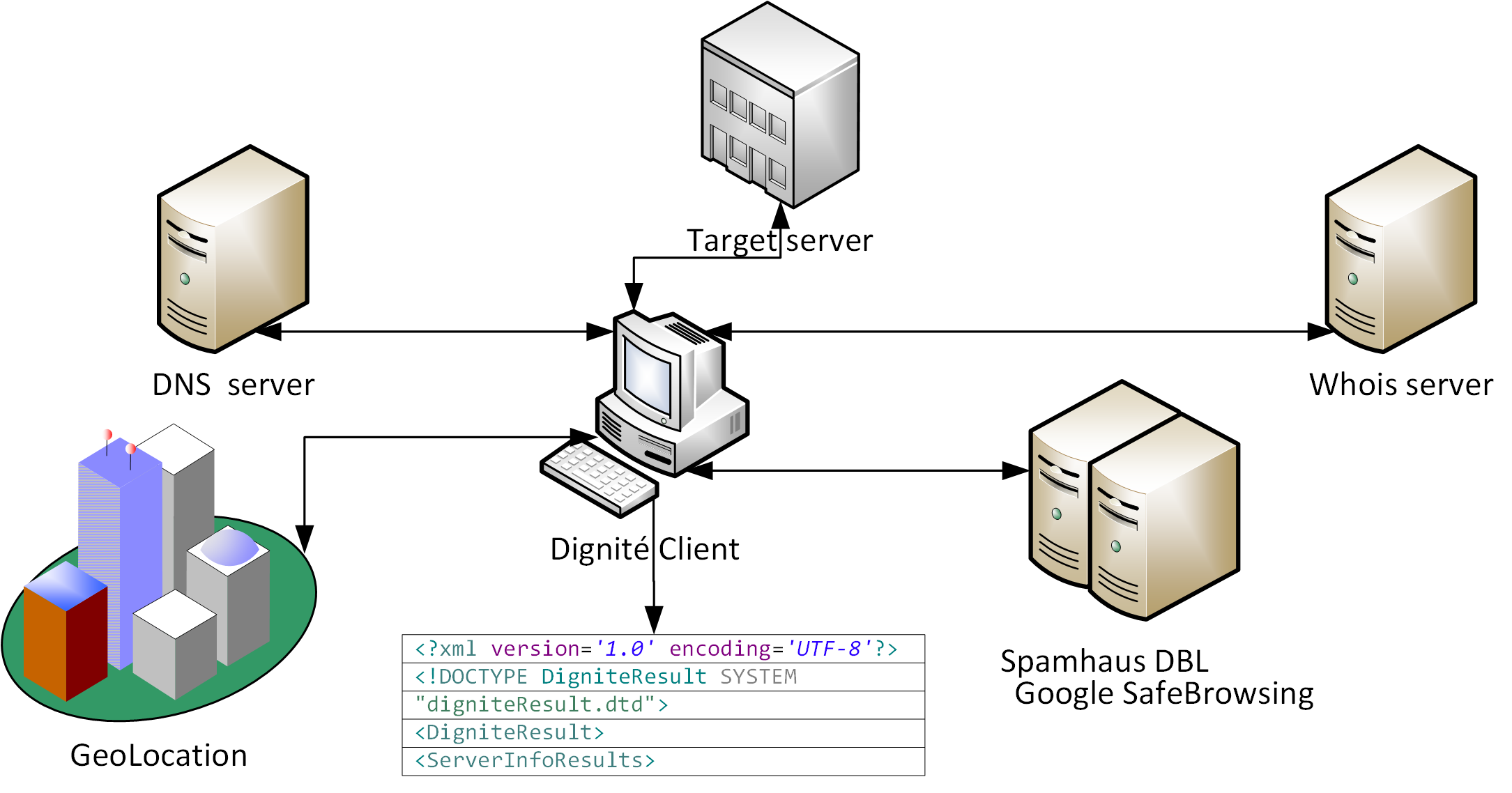}			
\end{textblock}

\begin{textblock}{154}(28,135)
	\begin{picture}(150,2)
		\put(0,0){\color{bfhgrey}\rule{150mm}{2mm}}
	\end{picture}
\end{textblock}
\color{black}

\begin{flushleft}

\vspace*{115mm}

\fontsize{26pt}{28pt}\selectfont 
\titel 				\\							
\vspace{2mm}

\fontsize{16pt}{24pt}\selectfont\vspace{0.3em}
\untertitel 				\\							
\vspace{5mm}

\fontsize{10pt}{12pt}\selectfont
\textbf{Bachelor Thesis} \\									
\vspace{3mm}

\begin{textblock}{150}(28,190)
\fontsize{10pt}{12pt}\selectfont
\end{textblock}

\begin{textblock}{150}(28,225)
\fontsize{10pt}{17pt}\selectfont
\begin{tabbing}
xxxxxxxxxxxxxxx\=xxxxxxxxxxxxxxxxxxxxxxxxxxxxxxxxxxxxxxxxxxxxxxx \kill
Branch:	\> Computer Science	\\			
Authors:		\> Thomas Ender, Patrick Vananti	\\					
Advisor:	\> Claude Fuhrer (Olivier Biberstein)		\\					
Expert:		\> Dr. Joachim Wolfgang Kaltz				\\					
Date:			\> \versiondate					\\		
\end{tabbing}

\end{textblock}
\end{flushleft}

\begin{textblock}{150}(28,280)
\noindent 
\color{bfhgrey}\fontsize{9pt}{10pt}\selectfont
Berner Fachhochschule | Haute école spécialisée bernoise | Bern University of Applied Sciences
\color{black}\selectfont
\end{textblock}

\end{titlepage}

%
%

%% file: vorspann/versionen.tex

\begin{textblock}{180}(15,150)
\color{black}
\begin{huge}
Versions
\end{huge}
\vspace{10mm}

\fontsize{10pt}{18pt}\selectfont
\begin{tabbing}
xxxxxxxxxxx\=xxxxxxxxxxxxxxx\=xxxxxxxxxxxxxx\=xxxxxxxxxxxxxxxxxxxxxxxxxxxxxxxxxxxxxxxxxxxxxxx \kill
Version	\> Date	\> Status		\> Remarks		\\
0.1	\> 18.02.2014	\> Draft		\> First version	\\	
0.2	\> 27.02.2014	\> Draft		\> Traces selected	\\	
0.3	\> 13.03.2014	\> Draft		\> API/library structure defined	\\	
0.4	\> 22.04.2014	\> Draft		\> Milestone: Base version	\\	
0.5	\> 08.05.2014	\> Draft		\> Milestone: Iteration 2	\\	
0.6	\> 05.06.2014	\> Draft		\> Milestone: Release Candidate	\\	
0.7	\> 07.06.2014	\> Draft		\> Anticipated review	\\	
0.8	\> 10.06.2014	\> Draft		\> Peer review	\\	
0.9	\> 12.06.2014	\> Draft		\> Ready for proof reading	\\	
1.0	\> 13.06.2014	\> Final	\>   	\\
\end{tabbing}

\end{textblock}

%% file: vorspann/management_summary.tex
\chapter*{Management summary}
\label{chap:managementSummary}

Web-based criminality like counterfeiting uses web applications which are 
hosted on web servers. Those servers contain a lot of information which 
can be used to identify the owner and other connected persons like hosters, 
shipping partners, money mules and more. 

These pieces of information reveal insights on the owner or provider of a 
fraud website, thus we can call them traces. These traces can then be used
by the police, law enforcement authorities or the legal representatives of 
the victim.

In our project 2 we had identified a vast range of possible traces. 
We had also considered their information content and existing limitations. 
During our Bachelor thesis, we have selected several traces and 
started the implementation of the API with its underlying library. 
After the successful implementation of the selected traces, we have created
a graphical user interface to allow the use of our solution without 
using a command-line interface.

To do so, we have learned to use the Scala Programming Language and its 
integration with Java code. The graphical user interface of our example application is
built using Scala Swing, the Scala adoption of the Swing Framework.
The test cases are defined using ScalaTest with FlatSpec and Matchers and executed using the JUnit Runner

%% file: kapitel/01-introduction.tex
\chapter{Introduction} 
\label{chap:intro}

This document contains the complete written documentation of the Bachelor thesis.
 
In the current, first, chapter, the project definition and administrative and planning aspects
are introduced. The following chapters deal with the traces we selected for implementation
as well as the implementation details regarding the \gls{API} and library.

Furthermore follow other important aspects including testing and performance and a short
explanation of the graphical user interface. After the conclusion, additional information is
presented in the annexes. As last annex, the documentation of the preliminary project 2 is 
provided for information purposes, as there are located the detailed descriptions of the 
traces we selected.

\section{Project definition}
\label{sec:intro_definition}
Many types of criminality use websites, hosted by web servers all-over the world, 
to disseminate illegal or fraudulent information. 
Such servers contain various pieces information, often called digital traces, 
which reveal insights on the type of illegal activity or on persons involved or related to the activity.

The goal of this Bachelor thesis is to develop modular and evolutive tools to acquire 
automatically a large range of digital traces. The various tools will be grouped together 
into an \gls{API} that could be easily integrated into larger applications. Moreover, an 
application that displays and generates a report of the results produced by the \gls{API} will be also developed.

Main objectives of the Bachelor thesis:
\begin{enumerate}
\item{Study the interesting digital traces and their limitations;}
\item{Study the Scala programming language that will be used for the implementation;}
\item{Design and implement a well structured \gls{API} for the extraction of the various traces;}
\item{Implementation of a graphical application that displays and reports the results of the \gls{API}.}
\end{enumerate}

The amount of digital traces that can be extracted by the \gls{API} will depend on the available time. 
The graphical application could display the screenshot of the website under extraction if time permits. 
The order of objectives corresponds to their respective priority during the project.
\newpage
\section{Deliverables}
\label{sec:intro_deliverables}
The essential documents to be delivered for this bachelor thesis are as follows:
\begin{itemize}
\item{Bachelor report, \\contains case study, explanation for the choices made, Scala concepts, \gls{API} structure, selected traces, structure and organization for this project.}
\item{Source code, \\contains all classes and the corresponding tests as well as the used build control files. All the classes and methods should be commented and for important parts of code an explanation is provided in the report. The relevant tests have to be done.}
\item{Graphical application}
\end{itemize}

The deadline for the delivery of the bachelor thesis is: 13.06.2014 17:00

\section{Used tools}
\label{sec:intro_tools}
\begin{itemize}
\item{Programming language: Scala 2.10.4 and Java 8}
\item{Unit testing: JUnit 4.11 and ScalaTest 2.10}
\item{Build control: Maven 3.2.1}
\item{Version control: Subversion repository https://svn.bfh.ch/repos/projects/dignite}
\item{Continuous integration: Jenkins \gls{CI}}
\item{Document editing: \LaTeX}
\item{Development Environment: eclipse 4.3.2 Kepler}
\end{itemize}

\section{Project planning}
\label{sec:intro_planning}

\begin{figure}[H] 
\center{\includegraphics[width=0.9\textheight, angle=90]{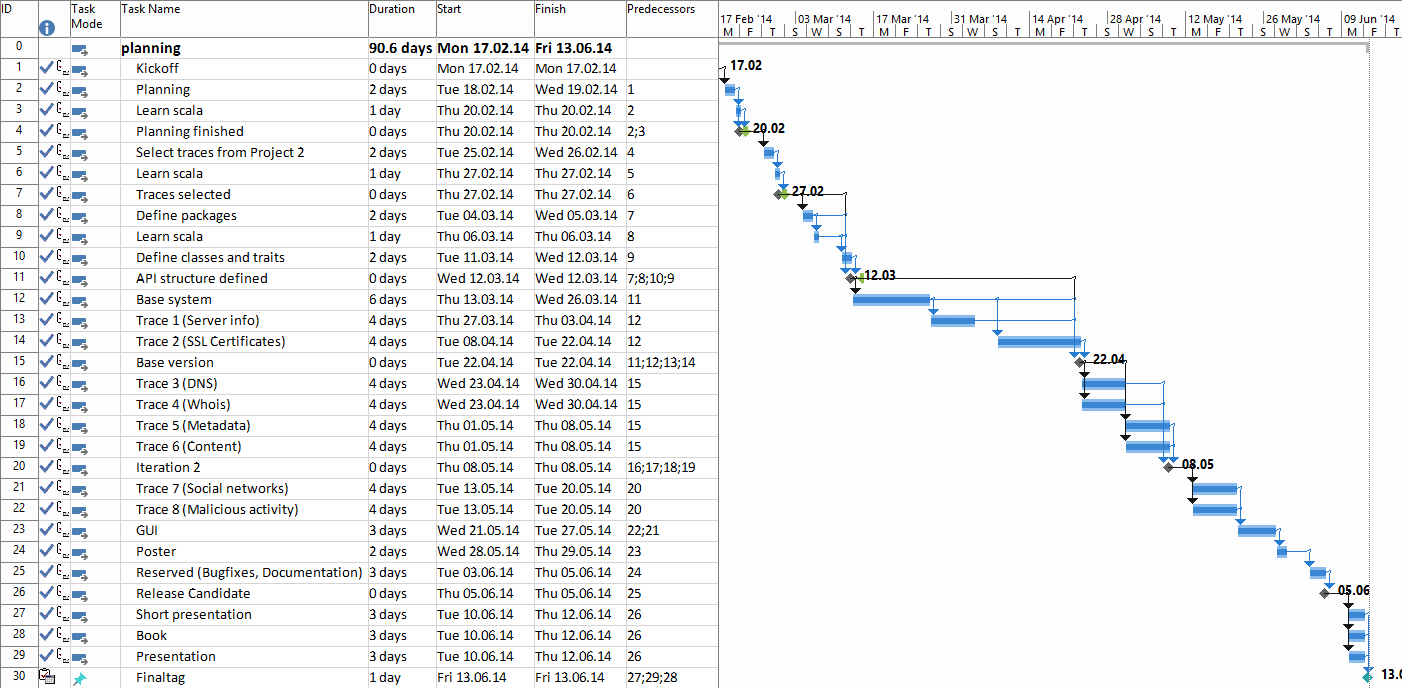}}
\caption{Gantt diagram}
\label{fig:gantt}
\end{figure}

\section{Project time-line}
\label{sec:intro_timeline}

Although the aforementioned Gantt diagram could imply a sequential, Waterfall-model
approach, the nature of this project demands for a more iterative, agile approach. 
Agile development processes like Scrum are based on iterations of fixed length, so-called
timeboxed Sprints. In this project, the three iterations ending with the milestones
"Base version", "Iteration 2" and "Release Candidate" are not strictly timeboxed but
are more or less 12 days long.

The overall structure can be seen on the following time-line:

\begin{figure}[H] 
\center{\includegraphics[width=0.9\textwidth]{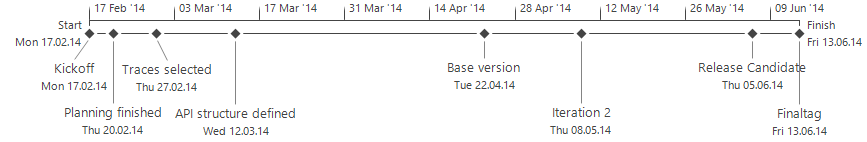}}
\caption{Time-line}
\label{fig:timeline}
\end{figure}

\section{Milestone analysis}
\label{sec:intro_milestones}

For each milestone, the current situation and the oncoming tasks need to be analyzed. 
Based on this analysis, a decision can then be taken whether the planning needs to be modified
or work can continue as planned.

\paragraph{Planning finished, Traces selected, \gls{API} structure defined} 
The first three milestones are grouped together as they are not linked to the implementation
and were already finished before the introduction of this analysis. 
But they have been reached as planned and no modification of the planning was necessary.

\paragraph{Base version}
This milestone is the first milestone directly connected to the implementation. Even though we had
some difficulties with the right structure for the trace results at first and a major rebuild was done for changing
to mainly asynchronous method calls, the milestone could be reached on time.
The planning was further refined to include an explicit time period for the graphical user interface.
Yet, no fundamental changes had to be done.

\paragraph{Iteration 2}
During Iteration 2, four traces were planned: \gls{DNS}, Whois, Metadata, Page Content.

While implementing the Whois trace, we encountered some problems with the format
of the obtained responses. Unfortunately, the Whois protocol does not specify a
result format. For this reason we can only do a best-effort parsing based on the 
most widely used format. To prevent information loss, the raw response is included
in the result as well. 

The Metadata trace uses regular expressions to filter the \verb|<meta>|-tags from the
\gls{HTML} data. The time used to define those expressions correctly was longer than expected. 
But at the end of this iteration we managed to extract the information successfully. For
the analysis of the \verb|robots.txt| file, we use regular expressions, where feasible.

Soon after starting the search for a web crawler to use with the Content trace, we 
noticed that, in fact, the Content trace and the Social Networks trace from the next 
iteration use the same mode of operation. They recursively visit a web site and 
check for some pattern in the body of the retrieved pages. Therefore we decided to 
combine those two traces into one, while using the time planned for both. This means that
traces 6 and 7 begin in Iteration 2 and finish implementation in the next iteration.

During this iteration, we also changed the package structure fundamentally, including new
sub-packages for the trace parameters and results as well as further separation of traits and their
respective implementations. This leads to a clearer structure and a better understanding of
the internal structure.

\paragraph{Release Candidate}
This last iteration of our project contains the remaining two traces Social Networks and Malicious Activity
as well as the \gls{GUI} and the final build configuration. During the planned time reserve, the jar building tasks
were included and bug fixes could be applied.

As described in the previous iteration, the Social Networks trace was integrated into the Page Content trace
and implemented as part of it. The Malicious Activity trace was relatively easy to implement, but needed to 
integrate an additional configuration option to provide the needed Google \gls{API} key, which has to be created by
every user at Google. For convenience, we deliver a sample configuration file, using our key, with the packaged files.

The \gls{GUI} development worked quite well, although the creation of the specific result panels used more time as 
we thought beforehand. Nevertheless we managed to have a working version including all traces in the graphical interface.
After finishing the main programming, we added the jar building step. This allows to create a single jar archive
containing all classes and dependencies.

Using this jar archive, we discovered some errors with unresolved resources, which were no longer found inside the jar.
But these could be fixed easily. Furthermore we started some performance measurements and added the build and launch 
instructions to the documentation. In the remaining time after this milestone until the end of the project the missing
parts of the documentation are completed and the various presentations and summaries written.

\section{Eclipse configuration}
\label{sec:intro_eclipseconfig}

To recreate the used development environment, the following additional software is needed
to be installed in an eclipse 4.3.2 Kepler installation:

\begin{itemize}
\item{Scala \gls{IDE} for eclipse, Update site: \url{http://download.scala-ide.org/sdk/e38/scala210/stable/site}}
\item{Scala Maven integration for eclipse, Update site: \url{http://alchim31.free.fr/m2e-scala/update-site}}
\end{itemize}

The needed libraries are defined in the Maven build control file \verb|pom.xml| and are automatically downloaded during import.

\section{Build project}
\label{sec:intro_build}

To build the library in an executable jar file, navigate to the \verb|dignite/library| folder 
and execute the following command:

\begin{lstlisting}[language={},keywords={},identifierstyle=\ttfamily,keywordstyle=\ttfamily]
> mvn clean package
\end{lstlisting}

This generates the \verb|dignite-<x.y.z>-complete.jar| in the \verb|target| subdirectory.

To build the \gls{GUI} in an executable jar file, change to the \verb|dignite/gui| folder
and execute the following commands:

\begin{lstlisting}[language={},keywords={},identifierstyle=\ttfamily,keywordstyle=\ttfamily]
> mvn org.apache.maven.plugins:maven-install-plugin:2.3.1:install-file 
-Dfile=<path-to>dignite-<x.y.z>-complete.jar 
-DgroupId=ch.bfh.ti.risis -DartifactId=dignite -Dversion=<x.y.z> -Dpackaging=jar 
-DlocalRepositoryPath=<workspace-path>\dignite\gui\repo

> mvn clean package
\end{lstlisting}

\newpage

The executable jar can then be started using the following command:
\begin{lstlisting}[language={},keywords={},identifierstyle=\ttfamily,keywordstyle=\ttfamily]
> java -Dconfig.resource=dignite.conf -jar dignite-gui-<x.y.z>-complete.jar
\end{lstlisting}

The same can be done with the library jar using the following command:
\begin{lstlisting}[language={},keywords={},identifierstyle=\ttfamily,keywordstyle=\ttfamily]
> java -Dconfig.resource=dignite.conf -jar dignite-<x.y.z>-complete.jar <target.tld>
\end{lstlisting}

In both cases the file \verb|dignite.conf| needs to be present in the same directory as the jar archive.
The file \verb|dignite.conf| should contain the following lines:
\begin{lstlisting}[language={},keywords={},identifierstyle=\ttfamily,keywordstyle=\ttfamily]
{
    "checkConnectionURL" : "www.ti.bfh.ch"
    "googleSafeBrowsingKey" : "AB<...>XA",
    "proxy" : {
        "host" : "",
        "port" : 0
    }
}
\end{lstlisting}

%% file: kapitel/02-selectedtraces.tex
\chapter{Selected traces} 
\label{chap:seltraces}

In the following section, we will present the traces we have selected and give a short reason for the decision. 
The following table gives an overview of the selected traces while noting the
corresponding number in the "\nameref{chap:projecttwo}" where the detailed description
can be found. The column "Trace X" corresponds to the trace number assigned in the section \nameref{sec:intro_planning}. 
\begin{table}[H]
\begin{tabular}{ c | l | c | l }
\hline
\textbf{\# in Project 2} & \textbf{Name} & \textbf{Trace X} & \textbf{Reason} \\ \hline
3.1.2. & Server \gls{IP} & 1 & needed for further actions \\ \hline
3.1.3. & Server info & 1 & server version and more \\ \hline
3.1.4. & \gls{SSL} certificate & 2 & gives certified information \\ \hline
3.2.1. & Forward \gls{DNS} & 3 & explore domain \\ \hline
3.2.2. & Reverse \gls{DNS} & 3 & discover hosting provider \\ \hline
3.2.3. & Whois & 4 & information about site owner \\ \hline
3.3.4. & Metadata & 5 & details about website creator \\ \hline
3.3.5. & robots.txt & 5 & hidden directories may be of interest \\ \hline
3.3.1. & E-Mail addresses and phone numbers & 6 & out-of-band connection to person \\ \hline
3.3.2. & External Hyperlinks & 6 & relationships to other sites \\ \hline
none & Find words in page & 6 & keyword categorization \\ \hline
3.4.1. & Google Analytics ID & 7 & relationships to other sites \\ \hline
3.7. & Social Networks & 7 & relationships to other sites \\ \hline
3.8.1. & Google Safe Browsing Score & 8 & verify malicious activity \\ \hline
3.8.2. & Spamhaus & 8 & verify malicious activity \\ \hline
\hline
\end{tabular}
\caption{Traces selected for implementation}
\label{tab:traces-table}
\end{table}

\section{Server \gls{IP} and info}
\label{sec:seltraces_ServerIP}

The presumably most common use case is the query using a known domain name. 
For many cases we need to have the corresponding \gls{IP} address. 
Therefore the retrieval of the \gls{IP} address is almost always the first step.
With the \gls{IP} it is possible to get the geographical location of the target. There are different levels of accuracy 
depending on which service was used to get that information. 
By retrieving server information like the operating system or the server software version, 
a further distinction between targets can be done and one could at least deduct a probability
whether or not two targets are owned by the same entity.

\section{\gls{SSL} certificate}
\label{sec:seltraces_SSLCert}

Nowadays, security is a very important aspect for every person or company on the Internet. Everyone tries to 
protect his information and his business, this also includes the criminal parts of the web.
More specific to \gls{SSL} certificates, the major goals are to encrypt the communication between two peers and to 
authenticate the server, so that nobody can impersonate the server.
Analysing the use of \gls{SSL} certificates reported by Ivan Ristic \cite{ristic:sslCertificateUsage}, 
36.52\% of all examined websites use \gls{SSL} certificates. 
This is one reason why those certificates are an important trace and the second reason is 
the quality of the information which is possible to acquire like holder name or company name.

\section{\gls{DNS}}
\label{sec:seltraces_DNS}

This is one of the few traces, which is almost guaranteed to exist. Without a working \gls{DNS}
configuration, a web site can not be accessed unless one knows the exact \gls{IP} address. 
Two ways of querying \gls{DNS} information are
possible. Forward queries obtain the corresponding \gls{IP} address to a given domain name,
while Reverse queries find the domain name connected to a given \gls{IP} address. 
Using forward enumeration, one can obtain all defined hostnames in a domain,
therefore hidden sites can be discovered. 
Reverse \gls{DNS} entries on the other hand are often done by and referring to the hosting provider, which can be
unveiled using a reverse lookup. This can be used if other sources are hidden or unavailable.

\section{Whois}
\label{sec:seltraces_Whois}

The whois database can be queried in two major ways. Using a domain-name, usually the registered owner and
technical contact should be listed. In more shady circles, this information is more and more protected by
specialized companies who replace the real entries with forwarders. 
The query based on an \gls{IP} address or address range yields the owner of this \gls{IP} range, who is registered at the \gls{IANA}.
This allows to obtain a rough geographical location of the target as well as a responsible contact. 

\section{Metadata / robots.txt}
\label{sec:seltraces_Metadata}

Using the meta-data included in the \gls{HTML} \verb|<head>| tag, the used tool or framework can be identified
more or less easily. Sometimes even the name of the web developer or its company are listed. Furthermore
common search-engine optimizations (\gls{SEO}) rely on keyword lists in the meta-data.
The \verb|robots.txt| file on the other hand can unveil hidden directories that might be rewarding when examined
later on.

\section{Search in content page}
\label{sec:seltraces_ContentPage}

Websites are containers for a lot of information, often people who manage web pages forget about
the importance and confidentiality of this data. 
The idea for this group of traces is to discover throughout the entire website the most useful information,
like thel relations with other websites or personal information such as email addresses and phone numbers.

\section{Social networks and Google Analytics}
\label{sec:seltraces_SocialNetworksAndVisitorStatistics}
Social networks are spread everywhere, people communicate with each other, they have relationships and friendships, 
they publish a lot of personal information so as to be recognized by other friends.
Social networks are at the same time a good platform for companies that want to do fast-spreading advertisement for free. 
Information found in social networks is not only within the network, but is also distributed on other websites, 
so you can more easily refer to one of them inside the network.
Google Analytics is a visitor statistics tool which is often used because it is free and shows a lot of statistics and data. 
These two types of traces give more possibility to identify the owner because they are widely used and 
a lot of personal information may be achieved.

\section{Malicious activity}
\label{sec:seltraces_MaliciousActivity}

Malicious Activity is not a trace like the other. It does not really reveal more information
about the website owner, nevertheless it is still useful to discover current or past
malicious activity using publicly available analysis tools like Google SafeBrowsing. 
Moreover it is possible to inform the user that the analyzed website is infected with malware.

%% file: kapitel/03-api.tex
\chapter{\gls{API} and library structure}
\label{chap:api}

\section{\gls{API} structure}
\label{sec:api_apistruct}

Before starting to explain our \gls{API}, it is necessary to understand what is an \gls{API} 
and what are the most important principles.
An \gls{API} is a set of programming objects or instructions which gives the possibility to 
access and use an already developed software. 
An \gls{API} developer releases its \gls{API} so other software developers can develop new applications using its existing work.
Nowadays most solutions for complicated problems, protocols and concepts are already developed, 
so a programmer does not need to write the entire code any more. Instead one can devote 
his time to solve the main part of the problem. Often users do not want to know how the 
software works inside but they simply want to use it without the risk of doing something wrong. 
Considering that \gls{API}s are intermediaries, they play an important rule to easy and good usability of a service.
Because of this reason, the key points to having a good \gls{API}, as found in \cite{blanchette:manualAPIdesign}, will be explained below.

\begin{itemize}
\item{\textbf{Easy to learn and memorize:} 
An easy to learn \gls{API} is composed of naming conventions, patterns and predictability. 
The concepts mentioned connect the programming experience with others already lived, so as to facilitate the learning process. 
This however does not mean that an \gls{API} must necessarily be minimalist, but the person who wants to use it, 
can easily start with little lines of code and may, with growing experience, take advantage of more complex functionality. 
The semantics should be simple, clear, and follow the principle of least surprise.}

\item{\textbf{Leads to readable code} 
Programming languages often allow writing compact code and using built-in facilities to solve a problem, 
but these can be a double-edged sword and the resulting code could be harder to read for an uninitiated programmer.
It is better to spend time writing a more verbose and clear code instead of saving time during the programming, 
as clear and concise code decreases the risk of mistakes. Nevertheless, this depends also on the programming language chosen.

A harder to read example:
\begin{lstlisting}
vehicle = new LandVehicle(Color.WHITE,2000,220,5.6)
\end{lstlisting}
A better readable example:
\begin{lstlisting}
landVehicle = new LandVehicle()
landVehicle.setColor(Color.WHITE)
landVehicle.setWeight(2000)
landVehicle.setMaxSpeed(220)
landVehicle.setConsumption(5.6)
\end{lstlisting}
}

\item{\textbf{Hard to misuse} 
A good-designed \gls{API} makes it easy to write correct code and reduces the possibility of,
intentionally or not, creating harmful applications.
This means to remove all those parts of the code where the user can interact directly with the internal logic of the \gls{API}.
Because such access can lead to errors, if you do not have the necessary knowledge.

Like in this example, where the user has the burden to provide the right index, which leads to mistakes:
\begin{lstlisting}
listOfElement.insertElement(0, "First Element")
listOfElement.insertElement(1, "Second Element")
listOfElement.insertElement(3, "Third Element") // Error
\end{lstlisting}
\newpage
This is a more concise example and the logic is managed by the backing list class:
\begin{lstlisting}
listOfElement += "First Element"
listOfElement += "Second Element"
listOfElement += "Third Element"
\end{lstlisting}
}
\item{\textbf{Easy to extend} 
Libraries grow over time. New features are added, consequently new classes must be developed and the existing classes 
may need to change some methods, parameters and enumerations. For this reason the \gls{API} should be designed with the concept 
of extensibility in mind. }

\item{\textbf{Complete} 
Every \gls{API} should allow users do everything they want, but in reality it is rarely so. 
Because of this, it should at least be possible for users to customize and extend the \gls{API}. 
Often libraries grow over time and at the same time the corresponding \gls{API} should be extended with the new features.}
\end{itemize}

In addition to the points mentioned above, we will use the \gls{SOLID} principles that help us design objects that are easy to maintain and use.
These principles were found in an article by Justin James\cite{solid:principles}. They have
been developed initially by Robert C. Martin known as "Uncle BOB".\cite{solid:unclebob}
\begin{itemize}
\item{\textbf{Single Responsibility Principle} states that every object should only have one reason to change, 
i.e. every object should perform one thing only.\\In our code we use this for example in the trace strategies and parameters which
are specific to their respective trace. In general we took care to separate the responsibilities as much as possible.}
\item{\textbf{Open-Closed Principle} states that classes should be open for extension and closed for modification.
\\All our classes allow sub-classing to extend the functionality. For example the \verb|MaliciousActivityStra-|\\\verb|tegyImpl| can be extended to 
selectively override the \verb|getGoogleSafeBrowsingScore(...)| or \verb|getSpam-|\\\verb|hausDBLScore(...)| method if desired.}
\item{\textbf{Liskov Substitution Principle} states that you should be able to use a derived class in place of a parent class and it must behave in 
the same manner.\\As we do not use subclasses in our Scala code, this principle is difficult to verify. In general, every instance of trait
might be replaced with its concrete implementation without negative effects.}
\item{\textbf{Interface Segregation Principle} states that clients should not be forced to depend on interfaces they do not use.\\
Considering the trace trait hierarchy, there is a common trait \verb|Trace| which is extended by specific traits for every trace 
like the \verb|WhoisTrace| or \verb|MetadataTrace| traits. Someone implementing a new metadata trace then only needs to implement 
the \verb|MetadataTrace| trait and not also all other trace methods defined separately in the other traits.}
\item{\textbf{Dependency Inversion Principle} helps to decouple your code by ensuring that you depend on abstractions rather than concrete implementations.\\
We took care to only use traits for our references unless not possible.}
\end{itemize}

\subsection{Packages and classes}
\label{subsec:api_apistruct_packages}
The design is a very important phase to be able to have a good, clear and simple \gls{API}, 
the points mentioned above are our base for the entire development.

The main package is \verb|ch.bfh.ti.risis.dignite.controller| where the fundamental classes for the \gls{API} reside.
To simplify the use we have defined the class \verb|TraceController| as main entrance point, 
it is responsible for different tasks.
As you can guess from the name of the class, the first task is allowing the user to query each 
type of trace and obtaining a trace result.
Other auxiliary tasks are also implemented, like verification and customization of an internet connection
or using an anonymized tunnel such as Java Anonymous Proxy or Vidalia.
 \newpage
The following figure shows the relation between the \gls{API} and its underlying library as well as
which packages are considered to belong to which part.

\begin{figure}[H] 
\center{\includegraphics[width=1.0\linewidth]{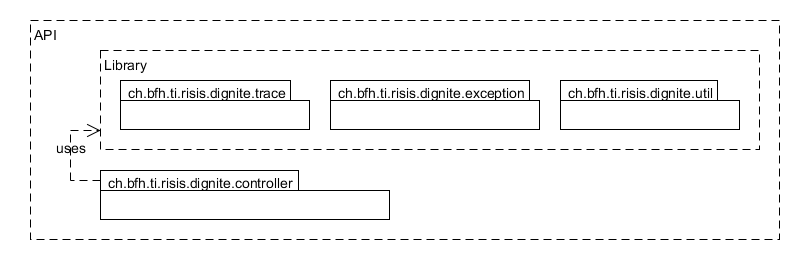}}
\caption{\gls{API} structure}
\label{fig:apistructure}
\end{figure}

\subsection{Use cases}
\label{subsec:api_apistruct_usecases}
Use cases are a method used during the design phase to identify and clarify system requirements.
A single use case is a sequence of interactions between a system and its users to reach a goal. 
The use case is divided in two main parts, the main success scenario for the case where everything works as planned
and the extensions where alternative interactions are explained in case something goes wrong during the main scenario.

Use cases are presented in two formats. The so-called "Brief format" describes 
the goal which should be achieved without detailing the sequence of actions. 
The "fully-dressed" use cases then elaborate the sequence of action in detail, 
while distinguishing between main success scenario and extensions.

As use cases deal with the interaction with the system, they need to describe users and other stakeholders. 
They are divided in primary, supporting and off-stage actors.
In our case, the primary actors are the \gls{API} users, not limited to but including human users. 
They can query different simple or combined traces and export it to another format. 
Supporting actors are all services that the \gls{API} uses to retrieve information:

\begin{itemize}
\item{\gls{DNS} servers}
\item{\gls{IP} \gls{API}}
\item{Google SafeBrowsing}
\item{Spamhaus Blocklist \gls{SBL}}
\item{Whois Server \verb|whois.iana.org|}
\item{Anonymization Provider (e.g. Vidalia)}
\item{the targeted server}
\end{itemize}

The following list shows the use cases of our \gls{API} in brief format. The detailed fully-dressed use cases 
and the derived System Sequence Diagrams (\gls{SSD}) and Sequence Diagrams (\gls{SD}) are placed in the appendix
"\nameref{chap:usecases-ssd}".
	\begin{itemize}
		\item\textbf{Use Case UC1:} Verify Internet Connection\\
		The API verifies the internet connection opening a communication channel with a reliable web server and returns the status.
		\item\textbf{Use Case UC2:} Configure Connection\\
		The user selects the connection parameters to use.
		\newpage
		\item\textbf{Use Case UC3:} Query a Digital Trace\\
		The user selects a type of trace and gives the necessary parameters, 
		then the system verifies the internet connection 
		and starts querying, until it is finished or until it times out.
		The system returns the trace result.
		\item\textbf{Use Case UC4:} Export traces \\
		The user selects the trace results, the path to export to and the export format,
		then the system verifies if the path exists and 
		then writes the trace result to the selected path with the chosen format.
	\end{itemize}	

\section{Library structure}
\label{sec:api_libstruct}

As a library is closely related to an \gls{API}, the same guidelines for a good design can be applied. 
But the most important characteristics of a library are its extensibility and the encapsulation. 
Although it has to be easily extensible, it should hide its internal structure as much as possible,
to facilitate its use. This leads to requiring encapsulation. 
 
The main goals of our library are:
\begin{itemize}
\item{easy extensibility}
\item{consistent result format}
\item{separation of traces and utility classes}
\end{itemize}

To achieve this, we use a main package \verb|ch.bfh.ti.risis.dignite| with 
sub-packages named \verb|controller|, \verb|exception|, \verb|trace| and \verb|util|. 
Inside the \verb|util| package, we place necessary helper classes. The trace acquisition is done using a specific
\verb|Trace| trait in the \verb|trace| package, with a corresponding \verb|TraceImpl| class in
the \verb|trace.impl| sub-package. The parameters are passed using a \verb|Param| class in the \verb|param| package
and the corresponding \verb|Result| class in the \verb|result| package is used to return 
the acquired data in a consistent format. The common methods and the public interface of those classes are defined in
two Traits named \verb|Trace| and \verb|Result|. Specific exception classes are placed in the package \verb|exception|. 
The point of entry for external classes is the trait \verb|TraceController| in the package \verb|controller|. 
\newpage
\begin{figure}[H] 
\center{\includegraphics[width=1.0\linewidth]{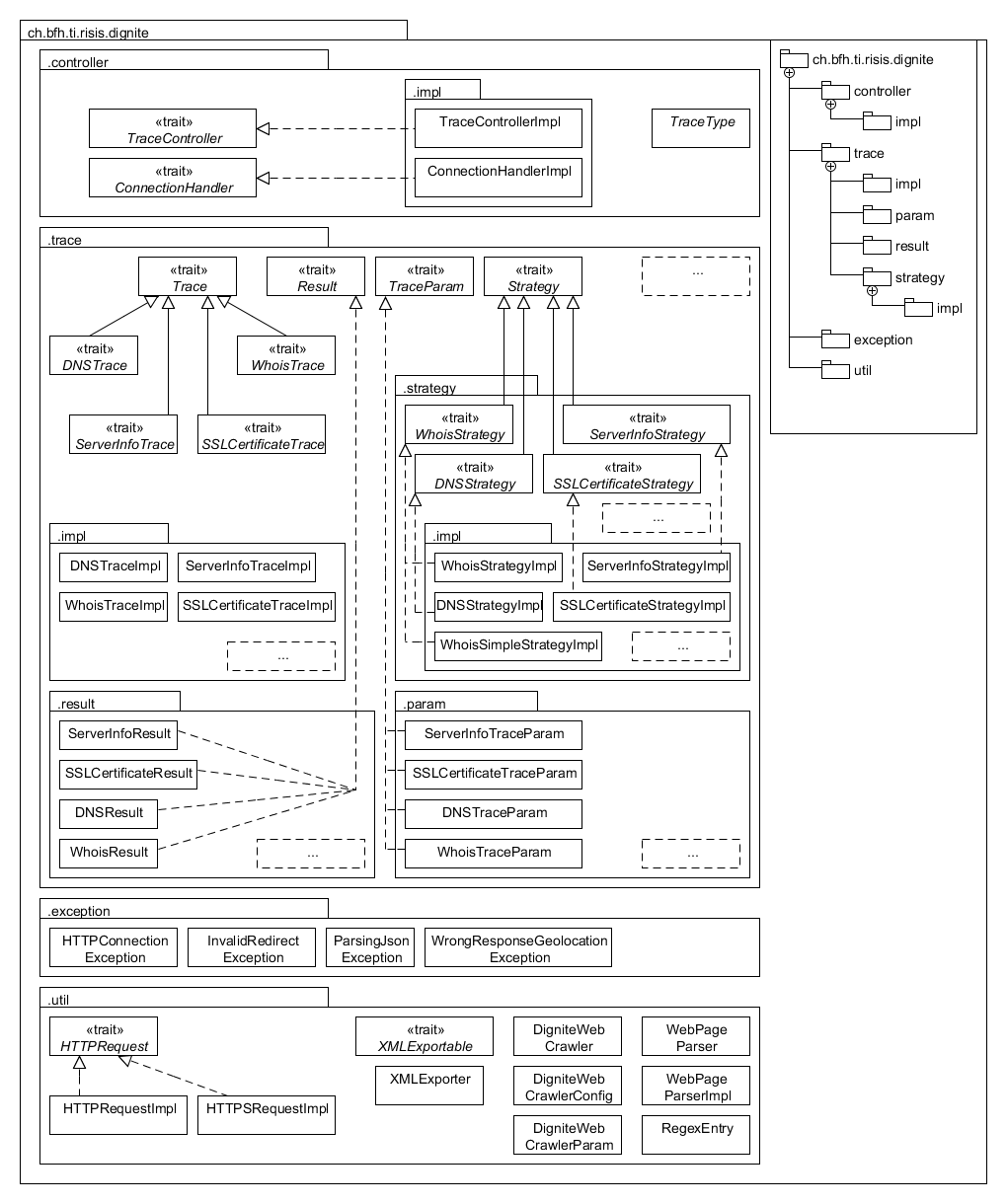}}
\caption{Library structure}
\label{fig:libstructure}
\end{figure}
\newpage
\subsection{Server info trace}
\label{subsec:api_libstruct_serverinfo}
In this and the following sub-sections, we will explain the general steps used in the processing of every trace. 
The following sequence diagram shows the handling of the server info trace. \\
\begin{figure}[H] 
\center{\includegraphics[width=1.0\linewidth]{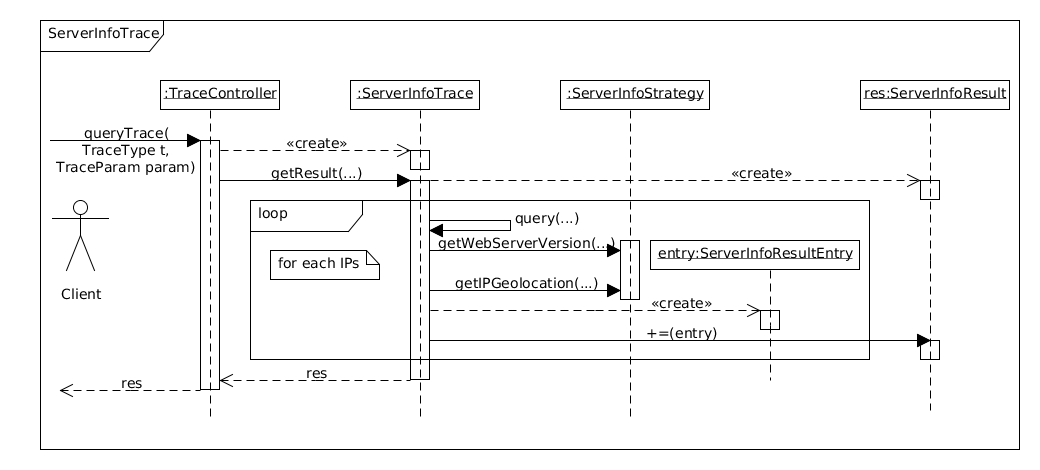}}
\caption{Sequence diagram of ServerInfoTrace handling}
\label{fig:seq-serverinfotrace}
\end{figure}

The current server info trace implementation allows users to acquire \gls{IP} addresses for the specified target, 
get their web server version and the geolocation. First of all, it gets all \gls{IP} addresses related with the target 
and then asynchronously computes the other information (web server version and \gls{IP} geolocation) for every address.
In the beginning was planned to measure the connection quality between application and the target, 
but Java does not provide libraries to qualify the connection. A possible solution would be the use of system commands. 
We have discarded the latter approach because it doesn't follow the platform independence principle of this project.

\paragraph{\gls{IP} addresses}

The implementation finds not only the \gls{IP} version 4 but also the new version 6, the trace will return all IPv6 addresses formatted as long.

\paragraph{Web server version}

Web servers can be divided in two different groups, in the first one the version 
can be discovered in the header of the server response.
In the second group unfortunately the version can not be discovered due to the hiding process made by the administrators. 
It removes also the version in the code of an error page. For this reason the strategy uses only information in the header, 
because if the administrator protects the version, then there is no way to find it elsewhere. 
The implementation works for both \gls{HTTP} and \gls{HTTPS}, it starts with \gls{HTTP} and if there is no response tries with \gls{HTTPS}. 
Certain websites allow only secure requests, so in this case it is also possible to retrieve the version.

\paragraph{\gls{IP} Geolocation}

For the \gls{IP} Geolocation different services are considered, like MaxMind, DB-IP, FreeGeoIP and IP-API.

Every one has positive and negative points, DB-IP offers a paid and free database to download, 
then the application does local queries directly on the database file. 
Unfortunately the information is updated only once a month and the free version is not very accurate.
The Maxmind service is practically the same thing as mentioned above, but with the added possibility of using a Web Service.

FreeGeoIP and IP-API do not have a local database, instead for every query an \gls{HTTP} request is sent to a specific page 
and the response contains the geolocation data. 
The requests have the format \verb|http://URL/<format>/<ip>| where format is one of \verb|json|, \verb|csv|, \verb|xml|
and for the \gls{IP} is given using the Dotted-Decimal notation\cite{wiki:dotdecimal}.
\newpage
The strong point of those external requests is, that every time a query is made, the most up-to-date 
result is obtained, obviously at the expense of increased traffic.

We know the Internet changes fast and regularly and even more for those sites with suspicious activity. 
For this reason it is better to have an updated database at every time.
We have chosen the IP-API service because it is more accurate, it is free and has less limitations.
The following listing shows an example request sent to the IP-API service:

\begin{lstlisting}[language={}]
HTTP request:
http://ip-api.com/json/46.126.85.7

HTTP response:
{
"status":"success",
"country":"Switzerland",
"countryCode":"CH",
"region":"05",
"regionName":"Bern",
"city":"Bienne",
"zip":"2504",
"lat":"47.1581",
"lon":"7.283",
"timezone":"Europe\/Zurich",
"isp":"Cablecom GmbH",
"org":"Cablecom GmbH",
"as":"AS6830 Liberty Global Operations B.V.",
"query":"46.126.85.7"
}
\end{lstlisting}

The usage is limited to 240 queries per minute, if this limit is exceeded the client is banned and
it is necessary to contact the service to get the \gls{IP} unbanned.

\subsection{\gls{SSL} certificate trace}
\label{subsec:api_libstruct_ssl}

\begin{figure}[H] 
\center{\includegraphics[width=1.0\linewidth]{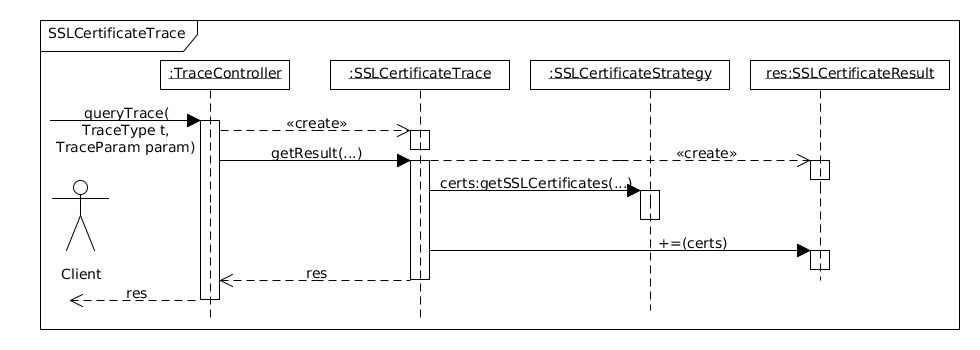}}
\caption{Sequence diagram of SSLCertificateTrace handling}
\label{fig:seq-sslcertificatetrace}
\end{figure}

The \gls{SSL} certificate trace uses existing code from the \verb|javax.net.ssl| package
to read the certificate chain. There is no problem for HTTPS websites conforming to 
the standards and using trusted certificates of acknowledged Certification Authorities.
For self-signed certificates or certificates from other Certification Authorities we need to disable
the integrated validity check performed by \verb|HttpsURLConnection|.
\newpage
To do this, we have to replace the \verb|X509TrustManager| used by the \verb|SSLSocketFactory|:

\begin{lstlisting}
/**
   * create a SocketFactory
   * ignoring whether a certificate is valid or not
   * WARNING: Use only for SSLCertificateTrace!
   * @return Trust-all SocketFactory
   */
  private def createSocketFactory = {
    // create custom trust manager to ignore trust paths
    val trm: TrustManager = new X509TrustManager() {
      def getAcceptedIssuers: Array[X509Certificate] = { null }

      def checkClientTrusted(certs: Array[X509Certificate], authType: String) {}

      def checkServerTrusted(certs: Array[X509Certificate], authType: String) {}
    };

    val sc: SSLContext = SSLContext.getInstance("SSL");
    sc.init(null, Array[TrustManager](trm), null);
    sc.getSocketFactory();
  }

\end{lstlisting}

\subsection{Whois trace}
\label{subsec:api_libstruct_whois}

\begin{figure}[H] 
\center{\includegraphics[width=1.0\linewidth]{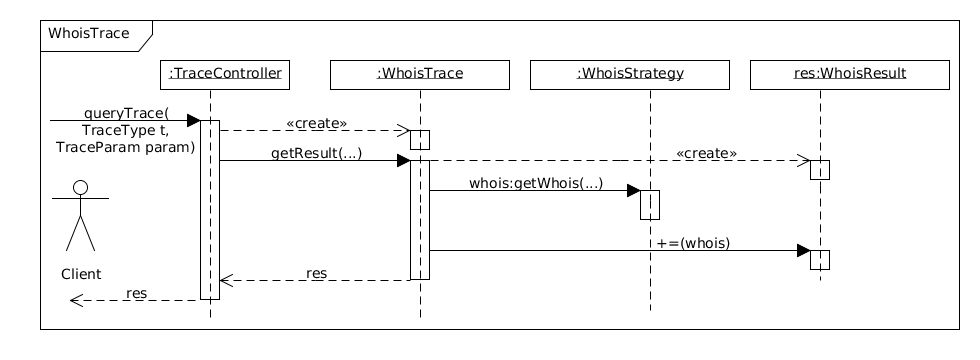}}
\caption{Sequence diagram of WhoisTrace handling}
\label{fig:seq-whoistrace}
\end{figure}

When implementing the Whois trace, we faced multiple difficulties. First of all, unlike \gls{DNS}, the Whois system
does not offer recursion. This means we cannot send a query to \verb|whois.nic.ch| when looking for information about
\verb|example.org|. As a consequence, the first step of a Whois resolution is to determine the Whois server to use. 
To achieve this, we send a first query to the Whois server at \verb|whois.iana.org| which sends 
back a \verb|refer: <whois server>| response if presented with a request it is not responsible for. 

After having found the right server to query, another problem lies in the nature of the Whois protocol. 
As defined in \gls{RFC}3912\cite{rfc:whois} the server's response has no defined formatting other than 
"the WHOIS server replies with text content". This makes it difficult to 
achieve a consistent result format with proper parsing.

On the other hand, more and more Whois servers start using a new pseudo-standard where
the lines respect the following format: \verb|Information name:  Information value| 

\newpage
To take advantage of this, we use the following 
regular expression to extract as much information as possible:

\begin{lstlisting}[language={}]
new Regex("""(?<tag>(?:\s|\w|\(|\)|\/)+):\s*(?<value>(?:\s|\w|\.|\,|\-|\+|\@)+)""")
//           ( named group "tag"       )    ( named group "value"             )
//           ( allows all characters,  )    ( allows all characters, spaces   )
//           ( spaces and '(' ')' '/'  )    ( and '.' ',' '-' '+' '@'         )
//           ( one or more times       )    ( one or more times               )
\end{lstlisting}

As a backup solution we still include the raw response in the resulting map, 
because we cannot guarantee that all information is properly
extracted from the server response.

There exist paid web services or paid libraries who manage the whole parsing of all 
different result formats, which could be used in a future version
when developing this work as a commercial product.

\subsection{\gls{DNS} trace}
\label{subsec:api_libstruct_dns}

\begin{figure}[H] 
\center{\includegraphics[width=1.0\linewidth]{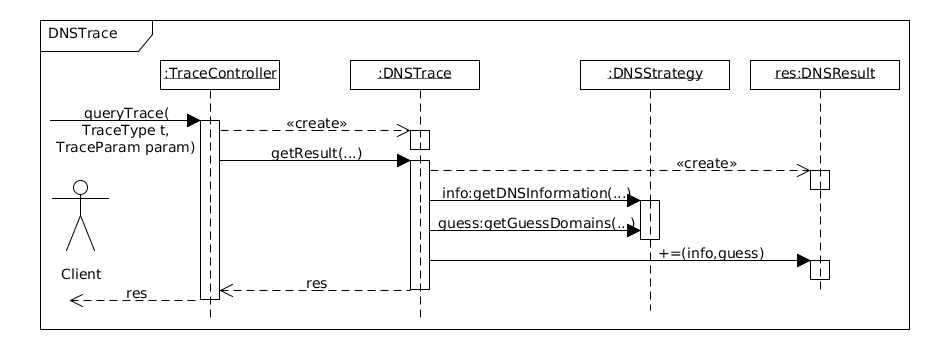}}
\caption{Sequence diagram of DNSTrace handling}
\label{fig:seq-dnstrace}
\end{figure}

Before explaining how this trace works, it is important to discuss the choice of the library.
The first library ( with classes \verb|DirContext|, \verb|Context|, \verb|Attributes|) which 
is present in the Java package tree since Java version 1.3, is a directory manager that can 
perform directory operations. This means it can not only execute DNS queries but also LDAP and 
other directory protocol queries.
The second library Dnsjava\cite{dns:dnsjava} is an implementation of \gls{DNS} in Java, it supports 
all defined records types and unknown types. It can authenticate \gls{TSIG} messages, verify a \gls{DNSSEC} 
message and transfer a zone.
After experimenting with the two procedures to get records, the second one is easier to use and 
returns a specific object depending on the type of record, so it is clearer how to get the 
specified information for every type.
Due to this advantage and the support of \gls{DNSSEC} the DNSJava library was choose.

The \gls{DNS} trace acquires all record types related to the target and tries to enumerate some hosts
by querying with a dictionary of common hosts.

First, it starts with making a forward query and getting all records. If a record is of type CNAME,
then it performs a recursive resolution until it receives a concrete type like A or AAAA. So the 
trace will return the entire information found with the chain of common names. The implementation 
has a recovery mechanism, so if there is a connection problem it will try again until the specified 
maximum number of attempts.

As second step, the \gls{DNS} trace reads the specified dictionary, and for each entry in the dictionary
it performs a query. If the response differs from NXDomain then the host exists. Only the existing hosts are returned.

The trace is capable to get AAAA records and IPv6 addresses.

\subsection{Metadata trace}
\label{subsec:api_libstruct_metadata}

\begin{figure}[H] 
\center{\includegraphics[width=1.0\linewidth]{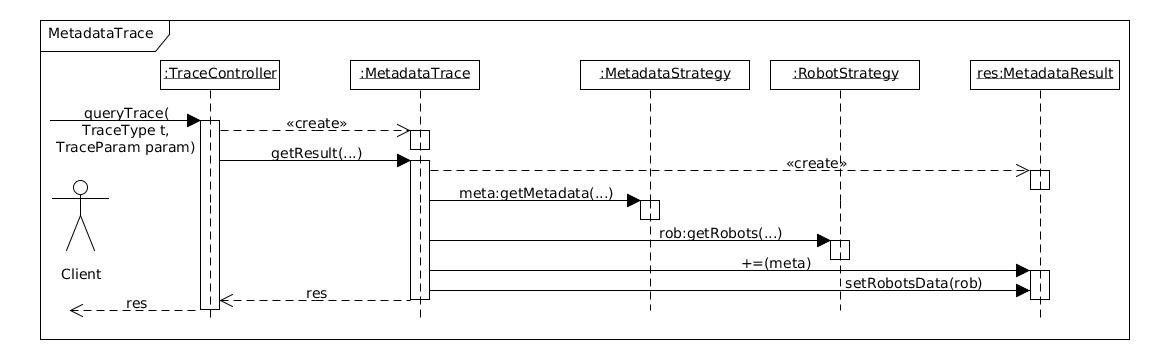}}
\caption{Sequence diagram of MetadataTrace handling}
\label{fig:seq-metadatatrace}
\end{figure}

The metadata trace extracts the \verb|<meta ...>| tags from the \gls{HTML} header. Those tags respect one of the following
three formats: 

\begin{lstlisting}[language={HTML}]
<meta charset="utf-8" />

<meta http-equiv="Content-Language" content="de-ch" />

<meta name="viewport" content="width=device-width, initial-scale=1, maximum-scale=1">

\end{lstlisting}

To find them in the \gls{HTML} header, the following regular expressions are used:

\begin{lstlisting}[language={}]
val metaCharsetFormat = new
 Regex("""<meta\s+(?<charset>charset)="(?<charsetContent>[^"]*)"\s*(?:\/)?>""")   
//                                     (group "charsetContent")
//                                     (allows all characters )
//                                     (different from "      )

val metaHttpEquivFormat = new
 Regex("""<meta\s+http\-equiv="(?<httpEquiv>[^"]*)"\s+content="                   
//                             (group "httpEquiv")
//                             (allows all characters )
//                             (different from "      )
                               (?<content>[^"]*)"\s*(?:\/)?>""")                
//                             (group "content")
//                             (allows all characters )
//                             (different from "      )

val metaTagFormat = new
 Regex("""<meta\s+name="(?<tagName>[^"]*)"\s+content="                            
//                      (group "tagName")
//                      (allows all characters )
//                      (different from "      )
                        (?<content>[^"]*)"\s*(?:\/)?>""")                       
//                      (group "content")
//                      (allows all characters )
//                      (different from "      )

\end{lstlisting}

The trace tries first to retrieve the information using a normal \gls{HTTP} GET request. 
If this should fail because the server uses \gls{HTTPS}, the request is re-issued using \gls{HTTPS}. 

In a second phase, the \verb|robots.txt| file is retrieved. Thanks to the wide-spread pseudo-standard, found 
on \cite{robots:nutshell}, the \verb|robots.txt| can easily be used line by line. 
First we planned using the following regular expressions:

\begin{lstlisting}[language={}]
val robotsDisallowEntry = new Regex("""Disallow:\s+(?<value>\S+)""")

val robotsAllowEntry = new Regex("""Allow:\s+(?<value>\S+)""")

val robotsUseragentEntry = new Regex("""User-agent:\s+(?<value>\S+)""")

val robotsSitemapEntry = new Regex("""Sitemap:\s+(?<value>\S+)""")

\end{lstlisting}

Although those regular expressions work well in several online regular expression test tools, we were unable 
to use them with the \verb|robots.txt| retrieved in this trace. 

We then changed to use the following expression extracting the type of entry and its
value at the same time. 

\begin{lstlisting}[language={}]
val robotPropertyFormat = new Regex("""(?<typer>[^\:]+)\:\s*(?<value>.*)""")
//                                     (group "typer" )     (group "value")
//                                     (allows all chars )  (allows all characters)
//                                     (different from : )

\end{lstlisting}

This allows to group entries of the same type together. If there
are entries or comments which do not respect this format, 
they are added with their respective line number.

\subsection{Page Content trace}
\label{subsec:api_libstruct_pagecontent}

\begin{figure}[H] 
\center{\includegraphics[width=1.0\linewidth]{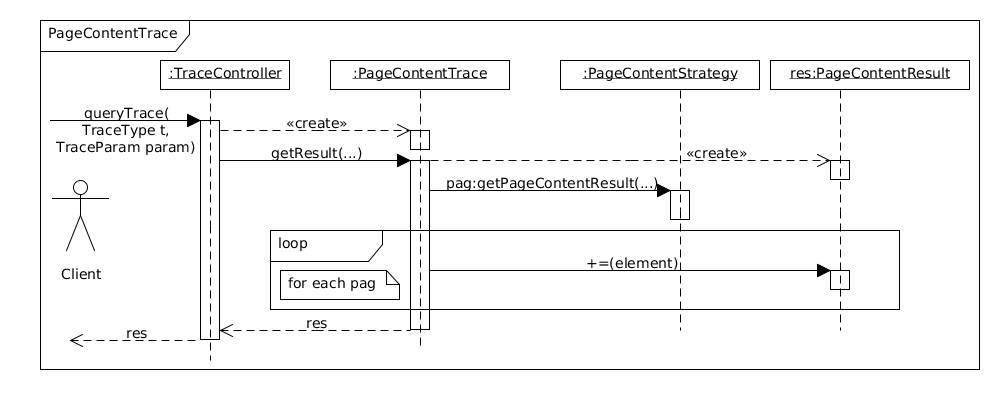}}
\caption{Sequence diagram of PageContentTrace handling}
\label{fig:seq-pagecontenttrace}
\end{figure}

The page content trace is a powerful tool for searching information not only in the specified target page 
but also in all related pages in the same domain. To achieve that the trace uses the web spider engine
Crawler4j\cite{page:crawler4j}, an open source Java crawler which provides a simple interface for crawling the Web.
The crawler implements almost all the logic except for a few methods in the class \verb|WebCrawler| that are 
specific to the usage and have to be implemented by a specialized class.
The trace class is obviously responsible to configure and start the crawler, then the \verb|DigniteWebCrawler| 
class extends the \verb|WebCrawler| and implements all remaining methods for the crawler engine, at last the class
\verb|WebPageParser| parses and saves the results. 

Due to the complexity and importance of implementation, in the next paragraphs will be explained 
how \verb|WebCrawler| and \verb|WebPageParser| are implemented.

As explained previously the \verb|WebCrawler| class has some methods that have to be implemented, 
the most important are \verb|shouldVisit| and \verb|visit|.

The first is the \verb|shouldVisit| method, it implements the algorithm that decides if a \gls{URL} found 
in a page should be visited by the crawler engine or not, this operation has to be done for each \gls{URL} found. 
The following code is the actual implementation and below there are more detailed explanations.
\begin{lstlisting}
  override def shouldVisit(url: WebURL): Boolean = {
    config.shouldVisitExtensionRegex findFirstIn url.getURL.toLowerCase match {
      case Some(url) => targetRegex findFirstIn url match {
        case Some(x) => true
        case None => {
          if (config.urlRelation) relations += url
          false}}
      case None => {
        if (config.imageRelation) {
          config.imageExtensionRegex findFirstIn url.getURL.toLowerCase match {
            case Some(url) => images += url
            case None =>}}
        false}}}
\end{lstlisting}

In the above code two regular expressions are used for the decision whether a \gls{URL} should be visited, 
the \verb|shouldVisitExtensionRegex| and the \verb|targetRegex|.
The \verb|shouldVisitExtensionRegex| allows only \gls{URL}s with a web page extension, or a \gls{URL} without 
extension like \verb|www.bfh.ch/welcome| because nowadays often mechanisms such as \verb|mod_rewrite|
are used which rewrite \verb|/welcome.html| to \verb|/welcome| or vice versa.
\begin{lstlisting}[language={}]
val shouldVisitExtensionRegex = new
 Regex(^(.*\/)+[^\.]*(\.(aspx?|cfm|chm|cms|(s|p|x|m)?html?|xhtml|jsp|mspx|
       php(3|4|5)?)(\?.*)?)?$)
\end{lstlisting}
The \verb|targetRegex| allows only \gls{URL}s in the same domain or any sub domain or the parent domain 
if the target has more then two domain levels. For example for the target \verb|www.bfh.ch| the allowed 
domains are \verb|www.bfh.ch|, \verb|*.www.bfh.ch| and \verb|*.bfh.ch|.
\begin{lstlisting}[language={}]
val targetRegex = new
 Regex((\w|\-|\.|\/|\:)*\.?bfh.ch\.?ch.*)
\end{lstlisting}
Knowing if a \gls{URL} should be visited is fundamental but it is also important to know which URLs do not match, 
because this information shows us the relations between other web sites or the image \gls{URL}s in a page.
To retrieve the relations just take all those \gls{URL}s that are web pages and do not match the domain rules. 
For the images take those \gls{URL}s that aren't web pages but instead match the image extension regular expression.
This is the \verb|imageExtensionRegex|.
\begin{lstlisting}[language={}]
val imageExtensionRegex = new
 Regex(^(.*\/)+[^\.]*(\.(jpe?g|png|gif|bmp|tif?f)(\?.*)?)?$)
\end{lstlisting}

The second important method is the \verb|visit| method that implements the logic for searching information 
in each page accepted by the above method.
The implementation calls our \verb|WebPageParser|, which parses the page using its regular expressions and adds the result to a Map.
\begin{lstlisting}
  override def visit(page: Page) {
    config.webPageParser.parsePage(page, parsedResultMap)
  }
\end{lstlisting}

The \verb|WebPageParser| contains the parse method called \verb|parsePage|, its implementation has a list of regular expression called \verb|RegexEntry|, 
where for each regular expression is checked if there are matches within the content of the page. When a match occurs, 
the result of the regular expression and the page \gls{URL} are added to the result map.
\newpage
Here is the implementation of \verb|parsePage|
\begin{lstlisting}
  def parsePage(page: Page, result: MultiMap[String, RegexResultEntry]) {
	// Other code not show for brevity
        regexList.foreach(regex => {
          if (!regex.getFirstFound || (regex.getFirstFound && !regex.isMatched))
            regex.getRegex findAllMatchIn data.getHtml foreach (textFound => {
              result addBinding (regex.getName, new RegexResultEntry(url, regex.getInterestingText(textFound.toString)))
              regex.matched
            }) })
	// Other code not show for brevity
\end{lstlisting}
Some more information about the regular expressions used. In our implementation there are two types, 
the first where the regular expression is executed for each page and the other which is executed only 
for the first occurrence. The second case improves the performance when the goal of a regular expression
is to find at least one occurrence. For example by searching a Google Analytics account,
it should be sufficient to find the first result, because for most cases only one Google Account 
is used for the entire web site.

Different aspects were taken into consideration before implementing the trace logic. 
Several crawlers have been analyzed and we chose to use Crawler4j, because it is a library based on Java,
open source and easy to integrate in our code. For parsing text we use regular expressions, 
which are one of the most powerful methods to parse a regular language. They allow sophisticated matches 
and their implementations are optimized and fast. In the above mentioned \verb|visit| method we decided
to call a parser class instead of implementing the code directly. This makes it easier to extend or change 
the implementation. To do this you need only to give the new parser implementation to the config object.
To avoid having search limitations due to a fixed implementation of search methods, we decides to implement 
a dynamic list in the \verb|WebPageParser|, where the user can add predefined regular expressions like GoogleAnalytics 
or SearchText or also a self written regular expression.

The trace can be parametrized with your optimizations, every usage needs some personalization to better fulfill 
the scope, for this reason we implement a specific class \verb|DigniteWebCrawlerConfig| which contains all 
configurations, like the regular expressions used or the specific crawler.

To better fulfill the tasks, Crawler4j implements multi-threading logic, it allows the crawl job to be 
divided in multiple subparts to speed up the operation. It is very efficient, it has been able to download 
and parse 200 pages per second on a Quad core PC with cable connection\cite{page:java-source}, it should 
easily scale to 20M pages.
The crawler follows the robot rules in the robot.txt file on the web site, according to the specification 
the crawler respects the Crawl-delay and Disallow rules if presents. The first rule defines what is the 
minimum time between each query, the second one determines which folder should not be crawled.

In this last paragraph are presented the different, already developed, \verb|RegexEntry|,
the first 5 \verb|RegexEntry| are added by default.

Email regular expression object used to get email address, even when the (AT) and (DOT) are used to protect the address from robots.
\begin{lstlisting}
object EmailRegex {
  val FindEmailRegex = new Regex("""[\w(\.|DOT|\(dot\)|\(DOT\))]+(@|AT|\(at\)|\(AT\))
  		[\w(\.|DOT|\(dot\)|\(DOT\))]+(\.|DOT|\(dot\)|\(DOT\))\w{2,4}""").unanchored
  def apply() = new RegexEntry ("Email", FindEmailRegex) 
}
\end{lstlisting}
Phone regular expression object used to get phone numbers with an international format like (+41 12 345 67 89, +41 320 123 45 65).
\begin{lstlisting}
object PhoneRegex {
  val FindPhoneRegex = new Regex("""\+\s*\d{1,3}[\s.()-]*(?:\d[\s.()-]*){10,12}(?:x\d*)?""").unanchored
  def apply() = new RegexEntry("Phone", FindPhoneRegex)
}
\end{lstlisting}
Google Web Analytics regular expression object used to get the account ID.
\begin{lstlisting}
object GoogleAnalyticsRegex {
  val FindGoogleAnalyticsRegex = new Regex("""\_gaq\.push\(\[\'\_setAccount\'\,\s*\'
  											(?<id>[\w\-\s]+)\'\]\)""").unanchored
  def apply() = new RegexEntry("GoogleAnalytics", FindGoogleAnalyticsRegex, true) {
    override def getInterestingText(textFound :String) = {
      textFound match {
      	case this.getRegex(id) => id
      }
    }
  }
}
\end{lstlisting}
Clicky Web Analytics regular expression object used to get the site ID and key
\begin{lstlisting}
object ClickyAnalyticsRegex {
  val FindClickyAnalyticsRegex = new Regex("""https?\:\/\/api\.clicky\.com\/api\/stats\/\d*\?
  								(?<id>site\_id\=[\d\-]*)\&(?<key>sitekey\=\d*)""").unanchored
    def apply() = new RegexEntry("ClickyAnalytics", FindClickyAnalyticsRegex, true) {
    override def getInterestingText(textFound :String) = {
      textFound match {
      	case this.getRegex(id,key) => id+" "+key
      }
    }
  }
}
\end{lstlisting}
Scripts regular expression object used to get the content of a \gls{HTML} <script> tag.
\begin{lstlisting}
object ScriptsRegex {
  val GetScript = new Regex("""<script[^>]*type=\"text\/javascript\"[^>]*(\/>|>.*
  					<\/script>)""").unanchored
  def apply() = new RegexEntry("Scripts", GetScript)
}
\end{lstlisting}
Facebook HTML 5 regular expression object used to get the Facebook information for
"Like", "Share page" or "Like page".
\begin{lstlisting}
object FacebookHTML5Regex {
  val FindFaceBookRegex = new Regex("""<div\s*class="fb-(send|post|follow|comments|
  						like-box|share-button|like)"[^>]*><\/div>""").unanchored
  def apply() = new RegexEntry("Facebook", FindFaceBookRegex)
}
\end{lstlisting}

Twitter regular expression used to get the twitter account.
\begin{lstlisting}
object TwitterRegex {
  val FindTwitterCounter = new Regex("""cdn\.api\.twitter\.com\/1\/urls\/count\.json\?url\=.*(\&|\&amp\;)
  									callback\=twttr\.receiveCount""").unanchored
  def apply() = new RegexEntry("Twitter", FindTwitterCounter)
}
\end{lstlisting}
\newpage
This object allows a user to search some text in each pages, it is possible to set case sensitivity.
\begin{lstlisting}
object SearchTextRegex {
  def apply(text: String, caseSensitive :Boolean = true) = {
    if(caseSensitive) new RegexEntry("Search[" + text + "]", new Regex(text).unanchored)
    else new RegexEntry("Search[" + text + "]", new Regex("(?i)"+text).unanchored)
  }
}
\end{lstlisting}

To add a new \verb|RegexEntry|, the user only needs to add an instance of it with the method \verb|addRegex|
of the \verb|PageContentStrategy| class. If someone wants to use the \verb|PageContentTrace| without the
pre-defined regular expressions, he can simply call the method \verb|removeAllRegexs| of \verb|PageContentStrategy|.
The following code explains how to create a \verb|PageContentTrace| without pre-defined \verb|RegexEntry|:
\begin{lstlisting}
  val pageContentStrategy = new PageContentStrategyImpl(conn)
  pageContentStrategy.removeAllRegexs
  val pageContentTrace = new PageContentTraceImpl(conn)
  pageContentTrace.setStrategy(pageContentStrategy)
\end{lstlisting}

\subsection{Malicious activity trace}
\label{subsec:api_libstruct_malicious}

\begin{figure}[H] 
\center{\includegraphics[width=1.0\linewidth]{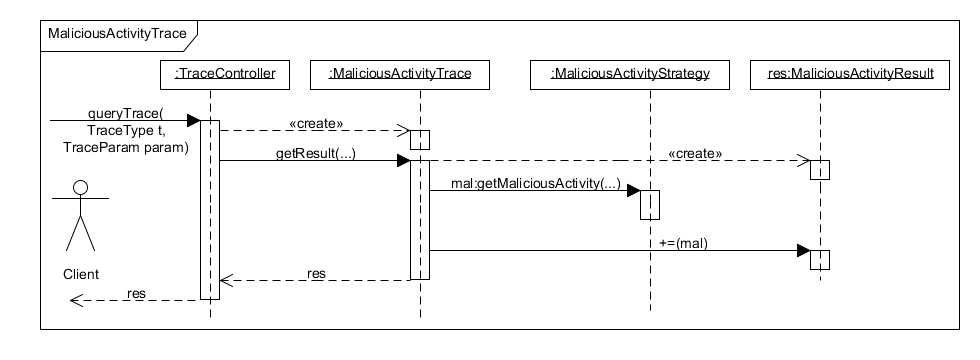}}
\caption{Sequence diagram of MaliciousActivityTrace handling}
\label{fig:seq-malicioustrace}
\end{figure}

In this trace the target address is used to query different external services, which provide a rating 
concerning malware or spam activity. The provided implementation of the MaliciousActivityStrategy uses the 
Google SafeBrowsing score which tracks malware and phishing activity and the Spamhaus Block List \gls{SBL} which tracks
spamming activity originating from the given target. 

To use the Google SafeBrowsing \gls{API}, a Google-issued \gls{API} key is necessary. The instructions how to obtain
such a key are found on the Google SafeBrowsing website \cite{license:google}:
\begin{quotation}
 In order to obtain an \gls{API} key, you must have a Google account. You may create a Google account or log in 
 with your existing Google account and sign up for the \gls{API} at \url{http://www.google.com/safebrowsing/key_signup.html}
\end{quotation}
Naturally, Google SafeBrowsing has limitations, but they are not very strict:

\begin{quotation}
We will limit the number of \gls{URL}s queried in a single POST request to be 500, which we believe is sufficient for most \gls{API} users. 
We will also limit the number of requests that can be made with a single \gls{API} key in a 24-hour period. If you expect to make more 
than 10,000 requests per day, you must contact us to have your \gls{API} key provisioned for additional users.
\end{quotation}

To retrieve the Google SafeBrowsing result for a given \gls{URL} the following \gls{HTTP} GET query is used. The \verb|<APIKEY>| has 
to be replaced with the users personal key and the \verb|<URL>| is replaced with the target \gls{URL}: 
\begin{lstlisting}[language={}]
https://sb-ssl.google.com/safebrowsing/api/lookup?client=dignite&apikey=<APIKEY>
 &appver=1.0&pver=3.0&url=<URL>
\end{lstlisting}
If the returned HTTP status code is 204, the given \gls{URL} is not known to the Google SafeBrowsing service, the status code 200 
indicates that the site is known to the service and the response body contains the type of activity detected which is either 
"phishing", "malware" or "phishing,malware". This result does not confirm that the site is currently infected or engaged in 
malicious activities but it was at least once in the past. 

The \gls{SBL} uses normal \gls{DNS} queries and the presence of a record in the special zone \verb|<target>.dbl.spamhaus.org.|
signifies that this domain has been reported as sender of Spam e-mails. A response of type "NXDOMAIN", meaning "Host not found",
confirms that there is no record for this target in the \gls{SBL}.
Usage of the \gls{SBL} is subject to three important limitations, otherwise the "Professional Use" paid service has to be used\cite{license:spamhaus}:
\begin{quotation}
Use of the Spamhaus DNSBLs via \gls{DNS} queries to our public \gls{DNSBL} servers is free of charge if you meet all three of the following criteria:\\
1) Your use of the Spamhaus \gls{DNSBL}s is non-commercial*,
    and\\
2) Your email traffic is less than 100,000 SMTP connections
    per day, and\\
3) Your \gls{DNSBL} query volume is less than 300,000 queries
    per day.\\
If you do not fit all three of these criteria then please do not use our public \gls{DNSBL} servers, instead see 'Professional Use'.
\end{quotation}

\subsection{Malicious relations trace}
\label{subsec:api_libstruct_combined}

Normally, every trace can be queried independently from each other, so one can give a domain name as input 
to the Whois trace and receives the Whois result in return. But sometimes one might wish to automatically 
query another trace with the output of the first trace. This combination of traces is wrapped in a
special case, the so-called "Combined trace".

In our library this approach is used for \verb|MaliciousRelationsTrace|.
It is composed of two strategies, the first one \verb|PageContentStrategy| is used to get external \gls{URL} relations.
The second one is the \verb|MaliciousActivityStrategy| which analyses each \gls{URL} found in the previous strategy. 

\begin{figure}[H] 
\center{\includegraphics[width=1.0\linewidth]{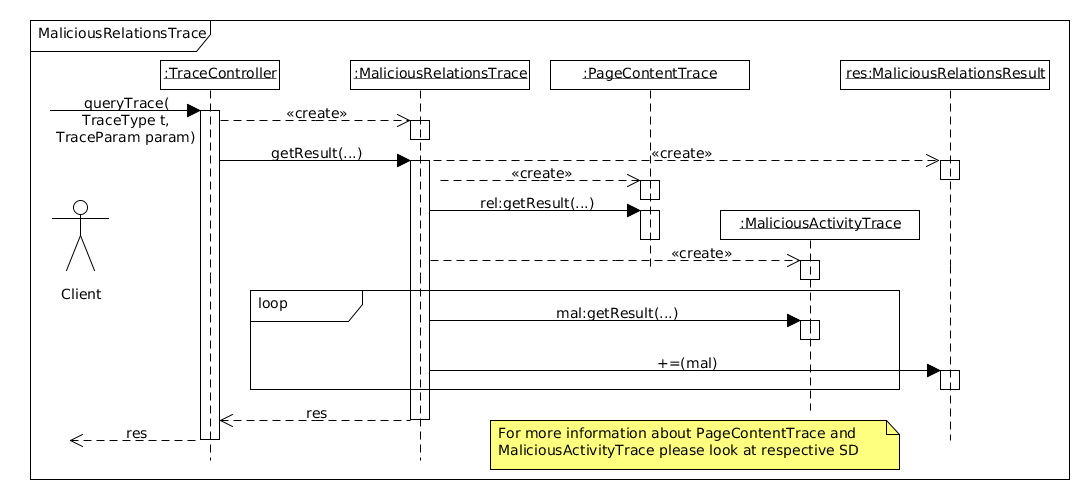}}
\caption{Sequence diagram of MaliciousRelationsTrace handling}
\label{fig:seq-maliciousrelationstrace}
\end{figure}

\section{Use and extend the \gls{API} and library}
\label{sec:api_useext}

In this part, we explain shortly how to use and how to extend our \gls{API} and library to fit custom needs.

To use the \gls{API}, a client can create an instance of the class \verb|TraceControllerImpl| in the package
\verb|ch.bfh.ti.risis.|\\\verb|dignite.controller.impl.TraceControllerImpl| and call its \verb|queryTrace(...)| methods.
As a guideline how to use the \verb|TraceController| the class \verb|DigniteLauncher| in the package 
\verb|ch.bfh.ti.risis.dignite| is provided. The \verb|DigniteLauncher| accepts one command-line argument 
which is the desired target. It then checks the connectivity and calls all traces implemented during this project, 
exporting the results to a \verb|result.xml| file.

To extend the \gls{API}, different approaches are possible. One way would be to further extend the existing
controller implementations in the \verb|ch.bfh.ti.risis.dignite.controller.impl| package and overwriting
or adding methods as needed. But we would recommend to either modify the traits defined in the 
\verb|ch.bfh.ti.risis.dignite.|\\\verb|controller| package directly, or to create an additional mix-in trait,
and then create new implementations using the modified traits.

To extend the library by adding a new trace, the \verb|ch.bfh.ti.risis.dignite.trace| package and its sub-packages
are the place to work at. The trait of the new trace should be defined directly in the \verb|ch.bfh.ti.risis.dignite.|\\\verb|trace| package.
In the sub-package \verb|ch.bfh.ti.risis.dignite.trace.impl| the trace implementation should be placed. 
The parameter and result classes should be defined in their respective \verb|ch.bfh.ti.risis.dignite.trace.|\\\verb|param| and 
\verb|ch.bfh.ti.risis.dignite.trace.result| sub-packages. 
Finally the strategy called by the trace implementation should be divided in a trait placed in the 
\verb|ch.bfh.ti.risis.dignite.trace.strategy| package
and an implementation found in the \verb|ch.bfh.ti.risis.dignite.trace.strategy.impl| package.

If only a new strategy to an existing trace should be added, this strategy needs to extend the 
corresponding strategy trait in the \verb|ch.bfh.ti.risis.dignite.trace.strategy| package and
its implementation is placed in the \verb|ch.bfh.ti.risis.dignite.trace.strategy.impl| package.

%% file: kapitel/04-securityandperformance.tex
\chapter{Testing and other aspects} 
\label{chap:secperf}

In the previous chapter, the \gls{API} and library structure have been discussed in detail. 
This chapter aims to explain important other topics related to our implementation that have not yet been discussed, but are still relevant for our project.
It will cover testing, performance and security aspects.

\section{Testing}
\label{sec:secperf_testing}

Before explaining our testing concept, it is important to understand the nature of our application.
Almost every class and object communicates over the Internet, directly or via external libraries like the \gls{DNS} trace.
The first idea that may come to mind is to test every trace with a real and external target, 
this method would allow us to test the entire code execution in a simple and efficient manner.
Unfortunately, we know that the Internet and websites are in constant change, for this reason it is 
very likely that the results obtained now are different from those obtained tomorrow. 
In addition to that some traces need a certain amount of time, for example the Page Content Trace can take 
from a minute to hours depending on the complexity of the targeted website.

Tests are an important part of our library, they are not only used to verify that our code works as designed 
during the first development, but also when someone will extend or improve some parts of the implementation. 
For this reason the tests should be stable over time and it should be fast and network independent to execute them.

So we decided to design our tests with mock objects, which are simulated objects that mimic the behavior of real 
objects in a controlled and predictable way. In our project, they will simulate typical service responses without 
using Internet connectivity. In most of our test cases we encounter that the real object would return
non-deterministic results, it may change its behavior over time and is slow to react. Therefore mock testing is to be
considered like described in the article "When To Use Mock Objects" \cite{page:mockobject}.

Together with mock objects we use the Scala Unit Testing called ScalaTest. This is a language specific test framework 
where the test definitions are more human readable, for more information see the \nameref{sec:scala_scalatest} appendix.

Testing all lines of code is not always possible, sometimes due to cost/time issues or due to the nature of the application.
In our case time and the nature of the application have played an important role in the development of tests.
We decided to concentrate our efforts, primarily by developing tests for the most important pieces of code, 
such as all strategies of each trace, secondly the rest of code where errors are likely and present a risk to correct
function.

To help qualify our tests we analyzed the code coverage, for that we used the Scala Code Coverage Tool scoverage,
this is a tool like Cobertura, but specific for the Scala language.

\begin{figure}[H] 
\center{\includegraphics[width=1.0\linewidth]{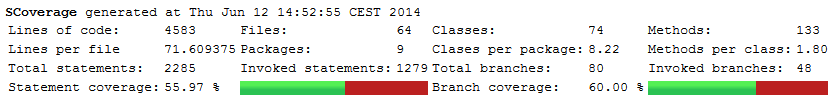}}
\caption{Scala Coverage}
\label{fig:scoverage}
\end{figure}

\newpage
Our degree of coverage is nearly 56\%, this result can be explained as follows:
\verb|toString| and \verb|toXML| methods are not tested, because we give priority to more critical pieces of code and they 
contain variables like the date which have unpredictable values.
\verb|XMLExportable| and  \verb|XMLExporter| classes are not tested because they use \verb|toXML| method which is sensible 
to the above problematic. \verb|DigniteLauncher| and  \verb|DigniteGuiLauncher| classes are not tested because they are only 
used to launch the queries, but are not a part of the library structure.

The biggest problem encountered during test development was to create a mock \gls{DNS} resolver. The desired behavior is simple, 
it sends a query and returns a \verb|Message| object containing the response.
For our tests, we wanted to return different responses generated by ourselves instead of using an Internet connection.
But to correctly instantiate a Message object, it is necessary to pass a byte array containing the \gls{DNS} response (RFC1034 
\cite{page:rfc1034}) received from the \gls{DNS} server. 
To generate this byte array correctly we decided to query a \gls{DNS} server and retrieved the response by sniffing the network
traffic with Wireshark. After capturing the complete response packet we select only the \gls{DNS} part of packet. This allows us 
to view the hexadecimal representation of the \gls{DNS} response, in this case for \verb|images.google.ch|:
\begin{lstlisting}
078e81800001 /* a....... */
0001000000000669/* .......i */
6d6167657306676f/* mages.go */
6f676c6502636800 /* ogle.ch. */
00ff0001c00c0005 /* ........ */
001000054540013 /* ....TT.. */
06696d6167657306 /* .images. */
676f6f676c650363 /* google.c */
6f6d00 /* om. */
\end{lstlisting}
This hexadecimal representation is then translated in a Scala Byte array and given to the Message constructor:
\begin{lstlisting}
new Message(Array(0x07.toByte, 0x8e.toByte, 0x81.toByte, 0x00.toByte, ..., 0x00.toByte))
\end{lstlisting}

After obtaining the correctly crafted Message object, it can be used as response for a specific mock call.
The tests took quite a bit of project time and in the end we are satisfied, the important pieces of code are tested
and the test results are stable.

\section{Performance}
\label{sec:secperf_performance}

Performance is another interesting part, used to verify the usability and efficiency of our library.
In this section we analyze the execution time of the implemented traces for some targets, located all-over the world. 
This is important for a significant performance measurement, because while using the product the target position is
not known in advance.

Targets have different sizes too, for example  \verb|google.com| or \verb|facebook.com| have a complex network structure 
with more \gls{IP} addresses, \gls{DNS} records and web pages than \verb|astab.ch|.
This allows us to know the performance for different target complexity as well.

As the number of queries and the latency to the target are an important part of the performance
measurements, we decided to measure this values using real-life targets and not using mock objects
like we did for testing. 

The following tables contain the results of the performance tests:\\\\\\\\
\begin{minipage}[H][5cm][b]{0,5\textwidth}
\centering
\begin{tabular}{l|l}
\textit{\textbf{Target}}   & \textit{\textbf{Time (seconds)}} \\ \hline
www.google.com             & 1.03                             \\
www.facebook.com           & 0.836                            \\
www.bfh.ch                 & 0.719                            \\
www.microsoft.com          & 1.857                            \\
www.apple.com              & 0.619                            \\
www.switch.ch              & 0.375                            \\
www.wikipedia.org          & 0.284                            \\
www.astab.ch               & 0.313                            \\
www.vananti.ch             & 1.2                              \\
www.openbsd.org            & 1.072                            \\ \hline
\textit{Average (seconds)} & \textit{\textbf{0.8305}}        
\end{tabular}
\captionof{table}{Performance ServerInfoTrace} \label{tab:serverInfoTraceTable} 
\end{minipage}
\begin{minipage}[H][5cm][b]{0,5\textwidth}
\centering
\begin{tabular}{l|l}
\textit{\textbf{Target}}   & \textit{\textbf{Time (seconds)}} \\ \hline
www.google.com             & 9.22                             \\
www.facebook.com           & 3.309                            \\
www.bfh.ch                 & 3.135                            \\
www.microsoft.com          & 4.049                            \\
www.apple.com              & 9.016                            \\
www.switch.ch              & 6.22                             \\
www.wikipedia.org          & 5.155                            \\
www.astab.ch               & 5.136                            \\
www.vananti.ch             & 5.979                            \\
www.openbsd.org            & 11.943                           \\ \hline
\textit{Average (seconds)} & \textit{\textbf{6.3162}}        
\end{tabular}
\captionof{table}{Performance DNSTrace} \label{tab:dnsTraceTable} 
\end{minipage}
\\\\\\\\\\
\begin{minipage}[H][5cm][b]{0,5\textwidth}
\centering
\begin{tabular}{l|l}
\textit{\textbf{Target}}   & \textit{\textbf{Time (seconds)}} \\ \hline
www.google.com             & 0.755                            \\
www.facebook.com           & 0.378                            \\
www.bfh.ch                 & 0.649                            \\
www.microsoft.com          & 1.596                            \\
www.apple.com              & 0.145                            \\
www.switch.ch              & 0.134                            \\
www.wikipedia.org          & 0.242                            \\
www.youtube.com            & 0.157                            \\
www.mozilla.org            & 1.603                            \\
www.twitter.com            & 1.233                            \\ \hline
\textit{Average (seconds)} & \textit{\textbf{0.6892}}        
\end{tabular}
\captionof{table}{Performance SSLCertificateTrace} \label{tab:sslCertificateTraceTable} 
\end{minipage}
\begin{minipage}[H][5cm][b]{0,5\textwidth}
\centering
\begin{tabular}{l|l}
\textit{\textbf{Target}}   & \textit{\textbf{Time (seconds)}} \\ \hline
www.google.com             & 1.655                            \\
www.facebook.com           & 1.053                            \\
www.bfh.ch                 & 0.564                            \\
www.microsoft.com          & 1.246                            \\
www.apple.com              & 2.335                            \\
www.switch.ch              & 0.649                            \\
www.wikipedia.org          & 1.489                            \\
www.youtube.com            & 1.069                            \\
www.mozilla.org            & 0.672                            \\
www.twitter.com            & 1.121                            \\ \hline
\textit{Average (seconds)} & \textit{\textbf{1.1853}}        
\end{tabular}
\captionof{table}{Performance WhoisTrace} \label{tab:whoisTraceTable} 
\end{minipage}
\\\\\\\\\\
\begin{minipage}[H][5cm][b]{0,5\textwidth}
\centering
\begin{tabular}{l|l}
\textit{\textbf{Target}}   & \textit{\textbf{Time (seconds)}} \\ \hline
www.google.com             & 0.46                             \\
www.facebook.com           & 1.071                            \\
www.bfh.ch                 & 1.218                            \\
www.microsoft.com          & 1.892                            \\
www.apple.com              & 0.141                            \\
www.switch.ch              & 0.134                            \\
www.wikipedia.org          & 0.897                            \\
www.astab.ch               & 0.643                            \\
www.vananti.ch             & 2.462                            \\
www.openbsd.org            & 0.86                             \\ \hline
\textit{Average (seconds)} & \textit{\textbf{0.9778}}        
\end{tabular}
\captionof{table}{Performance MetadataTrace} \label{tab:metadataTraceTable} 
\end{minipage}
\begin{minipage}[H][5cm][b]{0,5\textwidth}
\centering
\begin{tabular}{l|l}
\textit{\textbf{Target}}   & \textit{\textbf{Time (seconds)}} \\ \hline
www.google.com             & 0.675                             \\
www.facebook.com           & 0.403                            \\
www.bfh.ch                 & 0.135                            \\
www.microsoft.com          & 0.402                           \\
www.apple.com              & 0.212                            \\
www.switch.ch              & 0.348                             \\
www.wikipedia.org          & 0.154                            \\
www.astab.ch               & 0.17                            \\
www.vananti.ch             & 0.616                            \\
www.openbsd.org            & 0.728                           \\ \hline
\textit{Average (seconds)} & \textit{\textbf{0.3843}}        
\end{tabular}
\captionof{table}{Performance MaliciousActivityTrace} \label{tab:MaliciousActivityTraceTable} 
\end{minipage}
\\
\begin{minipage}[H][4cm][b]{0,5\textwidth}
\centering
\begin{tabular}{l|l}
\textit{\textbf{Target}}   & \textit{\textbf{Time (seconds)}} \\ \hline
www.astab.ch             & 61.2                             \\
www.benoist.ch           & 312.65                            \\
www.megawatch.co                 & 14400                            \\
www.rolexreplicati.com          & 325                           \\
www.replicageneve.com              & 162.3                           \\
www.vananti.ch             & 30.89                            \\
\end{tabular}
\captionof{table}{Performance PageContentTrace} \label{tab:pageContentTraceTable} 
\end{minipage}
\\\\Page Content Trace performance is directly proportional to the number of pages found and also to 
the number of \verb|RegexEntry| objects used to parse each page. For this reason the performances for 
this trace are not predictable. After analyzing the number of pages of a website and the time elapsed we 
have found an average of approximately 286 requests per minute.

\section{Anonymization}
\label{sec:secperf_anonymization}

When capturing large amounts of information from web sites with bad intent or not so legal contents, 
it can happen that they protect their resources against reconnaissance using \gls{IP} ban mechanisms or something similar. 
If this happens, it is necessary to change the \gls{IP} to launch the next query. Using an anonymization proxy allows users to easily 
switch their \gls{IP}. Anonymization is not only important for the above problematic but also for privacy, 
scanning malicious websites is not always appreciated by the owner. To avoid some repercussions 
it is recommended to use anonymization tools.

We tested our application with \gls{JAP} \cite{an:jap} anonymization tunnel, the most important limitation when using those 
services is the bandwidth which is very limited for the free version. Others tools can be used like Tor \cite{an:tor}.
To use any anonymization proxy, one only needs to set the \verb|Proxy| object to the controller class.

\section{Scaladoc}
\label{sec:secperf_scaladoc}
The comments for classes and methods in the code are formatted according to the \verb|scaladoc| standard. 
Every class and method is commented, complex methods contain additional inline comments.

This allows to generate a javadoc-like code documentation using the following maven command in the \verb|library| directory:
\begin{lstlisting}[language={}]
> mvn scala:doc
\end{lstlisting}
The generated documentation is then saved in the directory \verb|library/doc/scaladoc|

\section{Known bugs}
\label{sec:secperf_bugs}

The API and library in the current state have only two known bugs, all two of them are known Java bugs.

\verb|Proxy| class doesn't support \verb|HTTP_CONNECT| proxy, see JDK-6370908 \cite{bug:proxy} for more information.
To resolve this problem it is necessary to install the Java 8 environment.

\verb|HttpsURLConnection| class doesn't allow certificates which are signed with the \verb|MD2| algorithm and in that case 
returns an exception "Certificates does not conform to algorithm constraints". To resolve this limitation, it is necessary 
to re-enable this algorithm by editing the \verb|JDK_HOME/jre/lib/security/java.security| file and commenting out the following line 
\verb|jdk.certpath.disabledAlgorithms=MD2|. The solution was found in this article \cite{bug:ssl}.
Please pay attention to only use this for our library, because when using other applications like a Java browser it is very
important to keep this security setting activated, due to the vulnerabilities of the \gls{MD2}-based signature.
Our application doesn't attempt to verify the certificate signature, therefore it is not at risk for server impersonation attacks.

\section{Licenses}
\label{sec:secperf_licenses}

In our project two external libraries are used to implement trace strategy.
The \verb|dnsjava| library is used for every \gls{DNS} query and is distributed under the BSD license \cite{license:bsd}.
The \verb|crawler4j| library which is used for crawling a website in the Page Content trace, is instead distributed under the Apache 2.0 license \cite{license:apache2}.
Both licenses do not impose restrictions on the license used by the derived work.

%% file: kapitel/05-demoapp.tex
\chapter{Graphical application} 
\label{chap:demoapp}


Although we deliver a command-line based launcher (\verb|DigniteLauncher.scala|), 
a graphical user interface for simple use and a visually appealing presentation 
of the results was required. 

The graphical user interface should support the following actions:

\begin{itemize}
\item{Enter a target}
\item{Enter an \gls{HTTP} and \gls{HTTPS} target port}
\item{Select the desired trace}
\item{Start a query}
\item{Display the progress}
\item{Display the result}
\item{Export the results to a file}
\item{Change the configuration}
\end{itemize}

Based on this requirements two possible interface designs were produced as mock-ups. 
The difference between these two variants is mainly the position of the controls. 
In the vertical variant, all controls are grouped on the left side of the window, 
while the results are displayed on the right side. The horizontal version collates all
controls in the upper part of the window, next to the menu bar. 

\begin{figure}[H] 
\center{\includegraphics[width=1.0\linewidth]{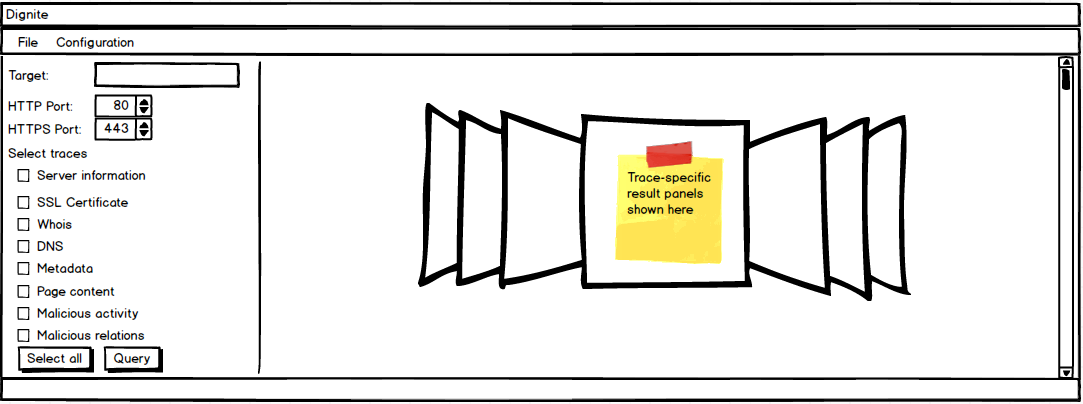}}
\caption{\gls{GUI}: Mockup with vertical orientation}
\label{fig:guimockvert}
\end{figure}

\begin{figure}[H] 
\center{\includegraphics[width=1.0\linewidth]{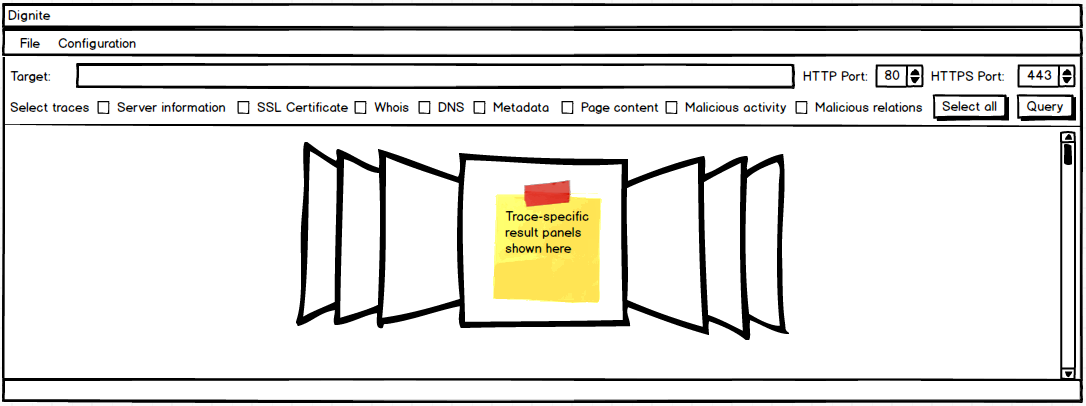}}
\caption{\gls{GUI}: Mockup with horizontal orientation}
\label{fig:guimockhoriz}
\end{figure}

At first glance, both variants seem equally apt to fulfill the requirements. But in the
case of a maximized or manually extended window, the vertical variant has a notable drawback.
The control block prevents that the list of results may use the whole width of the window. 
While using the horizontal variant allows to use all available screen space for the visualization
of the results, regardless of the number of result entries contained in the list.

Because of this, we created the "horizontal" user interface as shown on the next screenshot: 

\begin{figure}[H] 
\center{\includegraphics[width=1.0\linewidth]{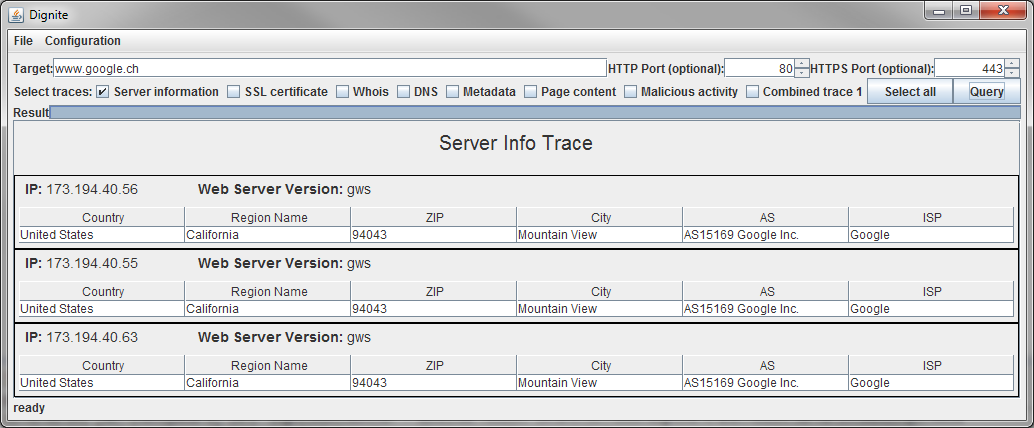}}
\caption{\gls{GUI}: Screenshot with Server information}
\label{fig:gui}
\end{figure}

The user interface uses standard Scala Swing components, only the result panels are custom-made 
sub-classes of the general Panel class to represent the collected information. The configuration
is loaded from a default configuration file and can be saved to a custom configuration file. For
the configuration handling, we use the Typesafe Config Framework \verb|com.typesafe.config|.

To start the graphical application, the class \verb|DigniteGuiLauncher| in the package \verb|ch.bfh.ti.risis.dignite.gui|
needs to be executed. No arguments are needed.

%% file: kapitel/06-conclusion.tex

\chapter{Conclusion and outlook} 
\label{chap:conclusion}


After the end of the programming phase of this project, we can assess our goals and determine where we see further potential.
The goals formulated in the project definition are the following:

\begin{enumerate}
\item{Study the interesting digital traces and their limitations;}
\item{Study the Scala programming language that will be used for the implementation;}
\item{Design and implement a well structured \gls{API} for the extraction of the various traces;}
\item{Implementation of a graphical application that displays and reports the results of the \gls{API}.}
\end{enumerate}

Those goals have all been reached and implemented in a good quality. Thanks to a detailed project planning and regular controlling
all milestones have been reached on time. The preliminary work in the project 2 during the last term allowed us to select the most
important and interesting traces we had identified. 
After the selection of traces we soon started the implementation of our \gls{API} and 
library, following a "learning by doing" principle. 

Thus we could learn how to use the Scala language while at the same time working
on our main goal, our \gls{API} and library. Although this sometimes caused bigger restructuring tasks, it was more informative, as we could
understand how to program using real life examples.

During the implementation of our \gls{API} and library we respected the \gls{SOLID} principles and the other points mentioned in the chapter
\nameref{chap:api}. 

The main difficulty we encountered during the implementation was to decide whether we created an asynchronous or
synchronous call structure inside our \gls{API} and library. Finally, we decided to leave it to the Client whether or not it would call our \gls{API}
in an asynchronous way and we implemented the synchronous call structure inside our library.

Another source of difficulties was the frequent interaction with Internet resources needed to achieve our goal. As long as everything goes according
to plan, it is similar to a local resource, but as soon as something goes wrong, the hassle begins. Unfortunately the error behavior is different
from resource to resource and this results in different errors and exceptions thrown by the Java networking stack. This makes it difficult 
to consider all possible exceptions. One of these special cases we encountered was the  "Certificates does not conform to algorithm constraints"
exception described in \nameref{sec:secperf_bugs}. Probably there are still exceptions in some special case we have not yet encountered. 

This limited predictability in the behavior of the queried resources led also to hindrances during the test development. But using mock objects 
for the testing, the network connectivity is not needed for the test execution and most difficulties can be avoided. On the other hand, the 
development of those mock objects was time consuming and had to be redone for every type of trace tested. For example we created a mock \verb|ConnectionHandler| 
to test the SSL certificate trace which returned the same set of three certificates every time. But for testing other traces like the Metadata trace
the \verb|ConnectionHandler| should behave differently. One particularly complex case was the mock for the DNS resolver, which is described in detail
in the section \nameref{sec:secperf_testing}.

The crawler library \verb|crawler4j| and its parametrization presented a challenge to correctly understand and use the crawler in our 
project with a suitable configuration that meets our needs. As the documentation did not really cover how to configure the
crawler and which classes had to be extended to override methods influencing the crawler's behavior, this information had to be extracted
from the examples provided and through experimental configurations.

Also the Scala integration for maven took some time to create a working version of the \verb|pom.xml| build control file.
 The Scala compilation for example needs to first convert the Scala code to Java code and then invokes the Java compiler 
 to create the corresponding bytecode. But after finding the correct maven plugin, this works well now.

\newpage
The last part of our project was to implement a graphical user interface to use our \gls{API}. During development of our library, we already
implemented a console based client, which calls the library in a synchronous way, therefore all traces are queried sequentially. The
graphical client on the other hand uses Futures to decouple the call from the result treatment and allows for faster response times 
to the user.

With the finished product, a user can now easily collect publicly available information about a given target website using only one tool
instead of many different command-line or graphical utilities. Furthermore all information retrieved is stored in a structured format 
using well-established \gls{XML} standards. This allows professional users like law enforcement as well as interested private users to better
understand the background of the target website and to base their decisions on concise information.

During the implementation we also found out about possible useful extensions of our product in the future which we did not consider during 
this project for time or compatibility reasons. One big task is the change from a Scala 2.10.x codebase to the new Scala 2.11.x codebase.
We could unfortunately not take this step, because of dependencies which did not exist for Scala 2.11 yet. 

Considering the upcoming and slowly growing support of \gls{DNSSEC} another valuable information would be the verification and analysis of its
special records and the zone signatures. Also implementing additional export formats based on the needs of the users are a future task
which can further improve the usability and utility mainly for professional users. This would allow to include the retrieved information
directly into other information management systems.

\paragraph{Personal conclusion}
We have enjoyed working on this topic which allowed us to apply the competences learned during our studies. During this thesis we could 
use the knowledge about networking and web applications we acquired during the network based modules, but never really applied during 
a project. Learning a completely new programming language gave us the opportunity to make a step forward and build up additional competences.

Working together as a team was a great experience as well. As we have similar interests and quickly understood in which direction we want
to go with our thesis we were off to a good start. Also throughout the project we got along well and could work effectively. After all
this time together showed us how important it is, to have a equally motivated and able partner, to continue even if the circumstances are
sometimes more difficult than at other times.

Last but not least, we would like to thank our expert, Dr Joachim Wolfgang Kaltz, and our advisors Claude Fuhrer and Olivier Biberstein
for their encouraging support and helpful advice.

%% file: vorspann/selbstaendigkeitserklaerung.tex
\chapter*{Declaration of conformity}
\label{chap:selbstaendigkeitserklaerung}

\vspace*{10mm} 

Ich bestätige mit meiner Unterschrift, dass ich meine Bachelor Thesis selbständig durchgeführt habe.
Alle Informationsquellen (Fachliteratur, Besprechungen mit Fachleuten, usw.), die wesentlich zu
meiner Arbeit beigetragen haben, sind in meinem Arbeitsbericht im Anhang vollständig aufgeführt.

\vspace{15mm}

\begin{tabbing}
xxxxxxxxxxxxxxxxxxxxxxxxx\=xxxxxxxxxxxxxxxxxxxxxxxxxxxxxx\=xxxxxxxxxxxxxxxxxxxxxxxxxxxxxx\kill
Ort, Datum:		\> Biel/Bienne, \versiondate \\ \\ 
Namen:	\> Thomas Ender 	\> Patrick Vananti \\ \\ \\ \\ 
Unterschriften:	\> ......................................\> ...................................... \\
\end{tabbing}

%% file: anhang/A0-usecases-ssd.tex

\chapter{Fully-dressed use cases and diagrams}
\label{chap:usecases-ssd}

\section{Fully dressed use cases}
\label{subsubsec:api_apistruct_usecases_full}
\begin{itemize}
\item\textbf{Use Case UC1:} Verify Internet Connection\\
\textbf{Primary Actor:} User\\
\textbf{Preconditions: } \\
\textbf{Main Success Scenario:}
	\begin{enumerate}
	\item User starts the verification.
	\item The system opens a http channel to a reliable web server.
	\item The system verifies the http channel.
	\item The system returns with an connected status.
	\end{enumerate}
\textbf{Extensions:}
	\begin{itemize}
	\item 2a: If the web server is not reachable.
		\begin{enumerate}
			\item The system tries to connect again for n times.
			\item The system returns with a disconnected status.
		\end{enumerate}
	\end{itemize}
\item\textbf{Use Case UC2:} Configure Connection\\
\textbf{Primary Actor:} User\\
\textbf{Main Success Scenario:}
	\begin{enumerate}
	\item User selects the connection's parameters.
	\item The system verifies the parameters.
	\item The system sets the parameters.
	\end{enumerate}
\textbf{Extensions:}
	\begin{itemize}
	\item 2a: If the connection's parameters are not valid.
		\begin{enumerate}
			\item The system returns with an exception.
		\end{enumerate}
	\end{itemize}	
\textbf{Postconditions:} User verifies the connection status \\
\item\textbf{Use Case UC3:} Query a Digital Trace\\
\textbf{Primary Actor:} User\\
\textbf{Preconditions:} User configured the connection (UC2)\\
\textbf{Main Success Scenario:}
	\begin{enumerate}
	\item User selects a type of traces.
	\item User selects the trace's parameters.
	\item The system verifies the trace's parameters.
	\item The system verifies the internet connection (UC1).
	\item The system starts acquiring the trace's information and stops.
	\item The system parses the retrieved information.
	\item The system returns a trace result.
	\end{enumerate}
\textbf{Extensions:}
	\begin{itemize}
	\item 3a: If the trace's parameters are not valid.
		\begin{enumerate}
			\item The system returns with an exception.
		\end{enumerate}
	\end{itemize}	
\item\textbf{Use Case UC4:} Export traces\\
\textbf{Primary Actor:} User\\
\textbf{Preconditions: } \\
\textbf{Main Success Scenario:}
	\begin{enumerate}
	\item User selects trace results to export.
	\item User selects exporting path.
	\item User selects the export format
	\item The system verifies the trace results are not empty or malformed.
	\item The system verifies that the path exists and it has writing permission.
	\item The system reads the trace result and formats with the right format.
	\item The system writes the file to the selected path.
	\item The system returns a OK status.
	\end{enumerate}
\textbf{Extensions:}
	\begin{itemize}
	\item 4a: If the trace results are empty or malformed.
		\begin{enumerate}
			\item The system returns with an exception.
		\end{enumerate}
	\item 4a: If the exporting path does not exists or the system has no writing permission
		\begin{enumerate}
			\item The system returns with a "path does not exists" exception or "no writing permission" exception.
		\end{enumerate}
	\end{itemize}		
\end{itemize}

\section{System Sequence Diagrams}
\label{subsec:api_apistruct_ssd}

\begin{figure}[H] 
\center{\includegraphics[width=1.0\linewidth]{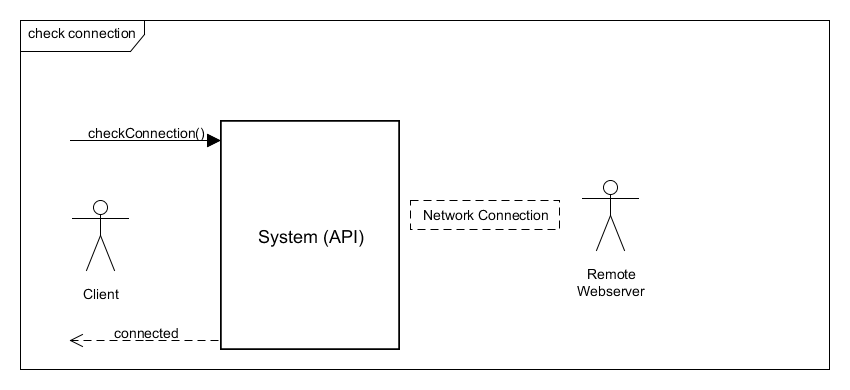}}
\caption{UC1: Verify Internet Connection \gls{SSD}}
\label{fig:ssd-01-checkconnection}
\end{figure}

\begin{figure}[H] 
\center{\includegraphics[width=1.0\linewidth]{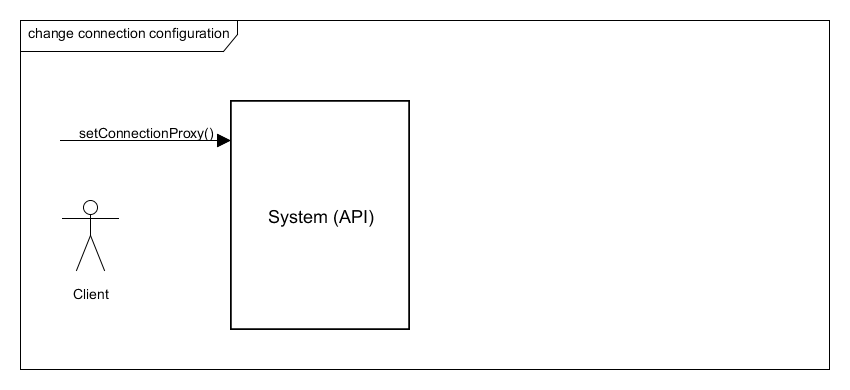}}
\caption{UC2: Configure Connection \gls{SSD}}
\label{fig:ssd-02-connectionsettings}
\end{figure}

\begin{figure}[H] 
\center{\includegraphics[width=1.0\linewidth]{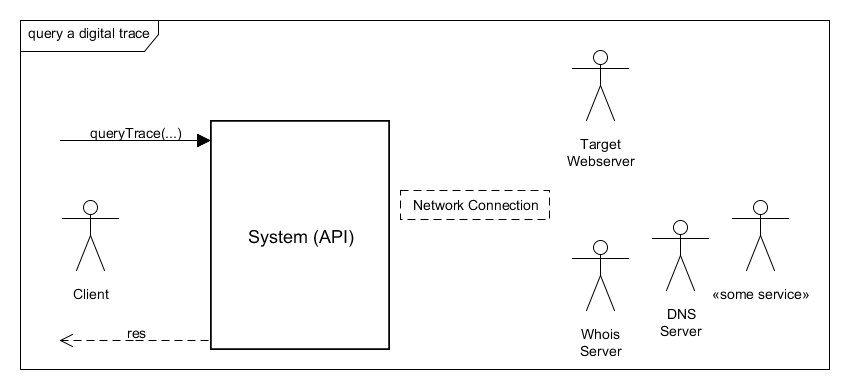}}
\caption{UC3: Query a Digital Trace \gls{SSD}}
\label{fig:ssd-03-tracehandling}
\end{figure}

\begin{figure}[H] 
\center{\includegraphics[width=1.0\linewidth]{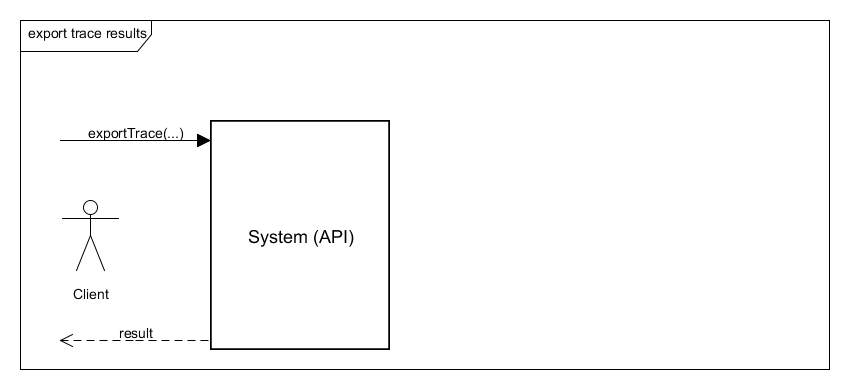}}
\caption{UC4: Export traces \gls{SSD}}
\label{fig:ssd-04-export}
\end{figure}

\section{Sequence Diagrams}
\label{subsec:api_apistruct_sd}

\begin{figure}[H] 
\center{\includegraphics[width=1.0\linewidth]{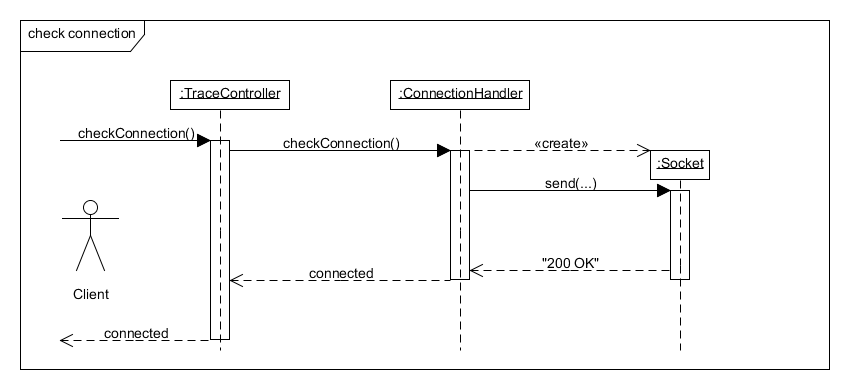}}
\caption{UC1: Verify Internet Connection \gls{SD}}
\label{fig:sd-01-checkconnection}
\end{figure}

\begin{figure}[H] 
\center{\includegraphics[width=1.0\linewidth]{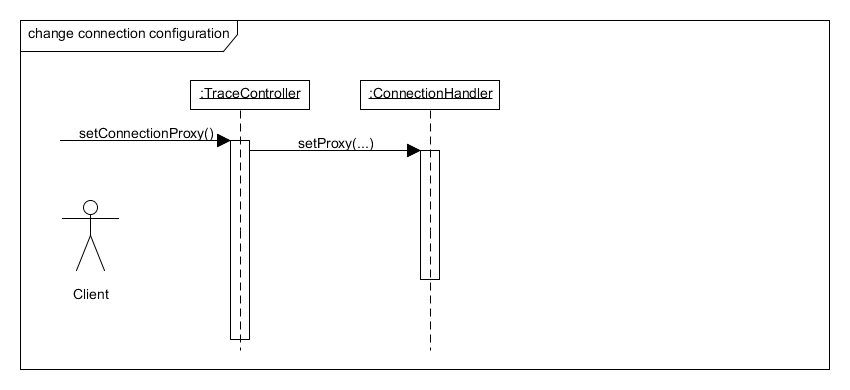}}
\caption{UC2: Configure Connection \gls{SD}}
\label{fig:sd-02-connectionsettings}
\end{figure}

\begin{figure}[H] 
\center{\includegraphics[width=1.0\linewidth]{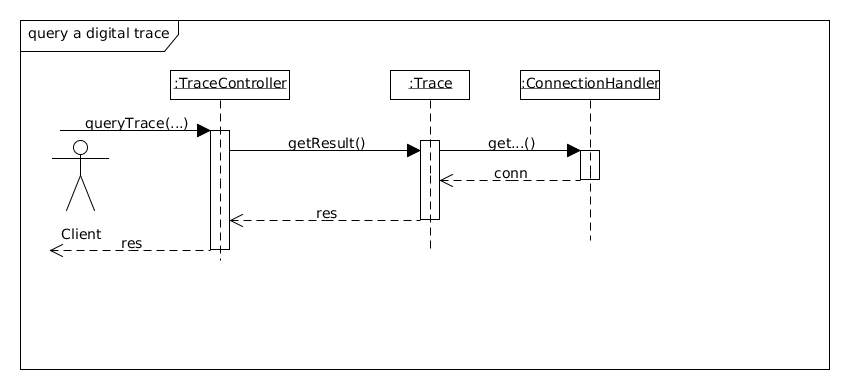}}
\caption{UC3: Query a Digital Trace \gls{SD}}
\label{fig:sd-03-tracehandling}
\end{figure}

\begin{figure}[H] 
\center{\includegraphics[width=1.0\linewidth]{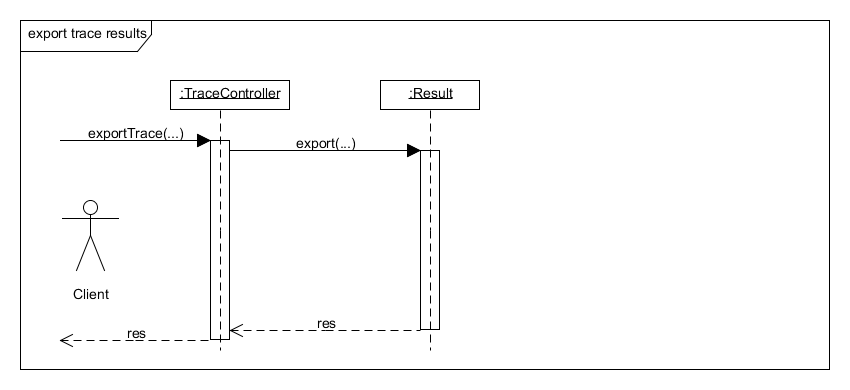}}
\caption{UC4: Export traces \gls{SD}}
\label{fig:sd-04-export}
\end{figure}

%% file: anhang/A3-scala.tex

\chapter{Scala concepts}
\label{chap:scala}

In this part, we will give a short introduction to the Scala Programming Language and 
explain the main concepts used in our \gls{API}. The goal is not to recreate a Scala tutorial 
but to highlight the key concepts and the relation to Java. This introductory part has been
taken from our Project 2 report \cite{endetvanap:projecttwo} and has been extended with 
more details about used concepts.

A typical Java-based \gls{API} is built using \textbf{interfaces} with \textbf{abstract classes} implementing them 
and additional classes as reference implementations or \textbf{helper classes}. There are not many restrictions 
to as how to build one's \gls{API}. But commonly all public methods are defined in interfaces and any application using 
the \gls{API} references only the interfaces. 

Although Scala is backwards compatible to Java, its internal structure is different. In the Scala Programming Language 
we differentiate between \textbf{Classes}, \textbf{Modules} and \textbf{Traits}. Scala also differs from Java in terms of 
syntax, primitive types (there are none) and other points. But for the sake of simplicity we will omit those here.
One important point in terms of file structure is, that in a single Scala source file multiple public classes can be defined.

A class is, like in Java, a definition of an object instantiated at runtime. 
Special types of classes like \verb|sealed| classes or "case classes" offer 
additional functionality compared to Java. 
A \verb|sealed| class is comparable to a \verb|final| class in Java, allowing subclassing only by classes defined in the same 
source file.
A "case class" is equivalent to a Java class which only public methods are accessors (or "getters"). It can be defined as follows:

\begin{lstlisting}
case class MyScalaUser(username:String, email:String)
\end{lstlisting}

A Scala module is best compared to a Java Singleton object, where only one instance may exist at runtime. 
It is defined by using the \verb|object| keyword. Scala Applications are also defined as modules.

\begin{lstlisting}
object Test extends Application {
  //...
}
\end{lstlisting}

Traits are the Scala equivalent of a Java interface but can also declare default implementations 
like an abstract class in Java would. 
A simple example:

\begin{lstlisting}
trait Similarity {
  def isSimilar(x: Any): Boolean
  def isNotSimilar(x: Any): Boolean = !isSimilar(x)
}
\end{lstlisting}

\section{Scala and multi threads application}
\label{sec:scala_actors}
A large number of applications needs more than one thread to operate. For example a Graphical User Interface which
interacts with an \gls{API}, or a simulation need to execute multiple methods simultaneously. Java natively allows the
manipulation and creation of threads, but the lock/shared memory principle is quite complex and difficult to master.
With bigger applications, it is complicated to identify the right variables and methods which need to be protected.
Scala uses another concurrency principle called Actors, they do not access the same memory but send and receive messages instead.

\subsection{Scala version < 2.10}
Actors are located in the default Scala package: \verb|scala.actors.Actor|

This is an example of a producer-consumer concurrency situation:

This part of code is the Consumer's implementation which waits for a Consume or Stop message. Note that no Producer
reference is saved in the class, because every received message contains the sender, which represents the Actor reference.
\begin{lstlisting}
import scala.actors.Actor

class Consumer extends Actor {
  def act() {
    var counter = 1
    loop {
      react {
        case Consume =>
          println("consumed product number " + counter)
          counter += 1
          sender ! Produce
        case Stop =>
          println("Consumer Stop")
          exit()
      }
    }
  }
}
\end{lstlisting}
The Producer class produces and then informs the consumer:
\begin{lstlisting}
import scala.actors
import scala.actors._

case object Produce
case object Consume
case object Stop

class Producer(quantity: Int, consumer: Actor) extends Actor {
  def act() {
    var counter = 1
    loop {
      react {
        case Produce =>
          // produce code
          println("produced product number " + counter)
          counter += 1
          consumer ! Consume
          if (counter >= quantity) {
            consumer ! Stop
            println("Producer Stop")
            exit()
          }
      }
    }
    
  }
}
\end{lstlisting}
\newpage
This is the starting object where the two Actors are started.
\begin{lstlisting}
object App {
  def main(args: Array[String]) {
    val consumer = new Consumer
    val producer = new Producer(10, consumer)
    println("Example of Producer Consumer with Actors")
    consumer.start
    producer.start
    producer ! Produce
  }
}
\end{lstlisting}
\subsection{Scala version >= 2.10}
From Scala version 2.10 onwards, the scala.actors package has been deprecated. 
Scala 2.10 uses the Akka Framework developed by Typesafe Inc. 
\cite{typesafe:akkaapi}
Actors are now located in the package: \verb|akka.actor.Actor|

Fortunately, this change happened before we started the implementation of our API/library 
and we can start directly with the latest version.

To illustrate the changes, the following example represents the same producer-consumer application as before.
This time implemented using the Akka Framework.

In the Akka Framework, the direct instantiation of Actors is normally not used. 
Instead an ActorSystem is created which groups together all Actors of an application. 
New Actor instances are then created using the factory method \verb|actorSystem.actorOf(Props properties, String name)|.
Moreover, the new Actor instances are automatically started by the ActorSystem. 
Explicit starting of an Actor instance is therefore not needed any more.

Although the Producer and Consumer classes still inherit from Actor, the application merely works with 
wrappers, the ActorRef objects. This fulfils one of Akka's key concepts, never to work with an Actor directly.

This is the rewritten starting object where the two Actors are created and started.
\begin{lstlisting}
object App extends Application {
	override def main(args: Array[String]) {
	    val system = ActorSystem("mySystem")
	    val consumer = system.actorOf(Props[Consumer], "myConsumer")
	    val producer = system.actorOf(Producer.props(10,consumer), "myProducer")
	    println("Example of Producer Consumer with Actors")
	}
}
\end{lstlisting}

As the instantiation of an Actor is done using the factory method mentioned earlier, 
a direct call to the Actors constructor is not possible any more. To achieve necessary 
initialization, the method \verb|preStart()| can be used. 

The Actor's work is done inside the 
\verb|receive = { ... }| part. 

The somewhat cumbersome \verb|act(){ loop{ react() {} } }| construct
from the former version is gone.
\begin{lstlisting}
import akka.actor
import akka.actor._

class Consumer extends Actor {
	// migrated using http://docs.scala-lang.org/overviews/core/actors-migration-guide.html
	private var counter = 0

	override def preStart() {
		// initialization code here
		counter = 0
	}
	
	def receive = {
		case Consume => 
			println("consumed product number" + counter)
			counter +=1
			sender() ! Produce
		case Stop =>
			println("Consumer Stop")
			exit()
	}
}
\end{lstlisting}
The Producer class produces and then informs the consumer. Here we need to pass parameters 
to the Producer's constructor. This is done by passing a Props object to the factory method.
To get this Props object and to call the constructor with it, the companion object of class Producer
is used to define a helper method \verb|props(quantity: Int,consumer:ActorRef): Props|. This
helper method then creates the appropriate Props object containing in fact a reference to an Actor
instantiated with the necessary parameters. 
\begin{lstlisting}
import akka.actor
import akka.actor._

case object Produce
case object Consume
case object Stop

object Producer {
  	/**
	 * Create Props for an actor of this type.
	 * @param quantity The quantity to be passed to this actor's constructor.
	 * @param consumer The consumer reference to be passed to this actor's constructor.
	 * @return a Props for creating this actor, which can then be further configured
	 * (e.g. calling '.withDispatcher()' on it)
	 */
	def props(quantity: Int,consumer:ActorRef): Props = Props(new Producer(quantity,consumer))
}
class Producer(quantity: Int, consumer: ActorRef) extends Actor {
	// migrated using http://docs.scala-lang.org/overviews/core/actors-migration-guide.html
	private var counter = 0

	override def preStart() {
	// initialization code here
	counter = 1
	consumer ! Consume
	}
	def receive() = {
		case Produce => {
			// produce code
			println("produced product number " + counter)
			counter += 1
			sender() ! Consume
			if (counter >= quantity) {
				sender() ! Stop
				exit()
			}
		}
	} 
}
\end{lstlisting}

\section{Scala and Swing}
\label{sec:scala_swing}
Swing is a famous framework used to create graphical interfaces in Java. Scala has imported and edited the framework 
using the Scala programming concepts so as to hide much of its complexity. The framework is located in the package: \verb|scala.swing|

Like in Java Swing, every object is a Component and all the classic components are also in the Scala library.

Scala simplifies their use by providing classes which already implement the main functionalities like application exit and start up.
One of this classes is the \verb|SimpleSwingApplication|.

Here is an example how to create a simple Swing application:
\begin{lstlisting}
    object ExampleSwing extends SimpleSwingApplication {
      def top() = new MainFrame {
        title = "Scala Frame"
        preferredSize = new Dimension { 
        	height = 100
        	width = 200
        }
        val button = new Button {
          text = "Scala+"
        }
        val label = new Label {
          text = "Welcome"
        }
        contents = new BorderPanel{
          layout(label) = BorderPanel.Position.North
          layout(button) = BorderPanel.Position.South
        }        
      }
    }
\end{lstlisting}

\begin{figure}[h]
\begin{center}
\includegraphics[scale=0.6]{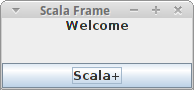}
\caption{Swing application with one button and one label}
\end{center}
\end{figure}
Every graphical interface can be decomposed in different subcomponents, in Swing these components are: one Frame and many Panels.
Every Panel has a layout and it may contain several other components, instead the Frame is the main component and has at most one child.
\begin{lstlisting}
    object ExampleSwing extends SimpleSwingApplication {
      def top() = new MainFrame {
        val menuPanel = new BoxPanel(Orientation.Vertical) {
          contents += new Label("Menu1")
          contents += new Label("Menu2")
          border = Swing.LineBorder(new scala.swing.Color(0))
        }
        val mainPanel = new BoxPanel(Orientation.Vertical) {
          contents += new Label("main text")
          contents += new Button("Button 1")
          border = Swing.EmptyBorder(10)
        }
        contents = new BorderPanel{
          layout(menuPanel) = BorderPanel.Position.West
          layout(mainPanel) = BorderPanel.Position.Center
        }        
      }
    }
\end{lstlisting}
\begin{figure}[h]
\begin{center}
\includegraphics[scale=0.6]{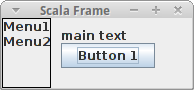}
\caption{Swing application with more panels}
\end{center}
\end{figure}

Handling events is a very important task when working with interactive graphical interfaces. 
For example an event occurs, when a button is clicked. In Scala Swing there are two possibilities, 
one is using the Reaction principle and the other is working with Actions. Here is an example using Reaction.
\begin{lstlisting}
    object ExampleSwing extends SimpleSwingApplication {
      def top() = new MainFrame {
        val lightOnText = "Light ON"
        val lightOffText = "Light OFF"
        val button = new Button {
          text = "Light ON/OFF"
        }
        val label = new Label {
          text = lightOffText
          listenTo(button)
          reactions += {
            case e: ButtonClicked => text = if(text==lightOffText) lightOnText else lightOffText
          }
        }
        contents = new BorderPanel{
          layout(label) = BorderPanel.Position.North
          layout(button) = BorderPanel.Position.South
        }        
      }
    }
\end{lstlisting}
Here is an example using Action
 \begin{lstlisting}
    object ExampleSwing extends SimpleSwingApplication {
      def top() = new MainFrame {
        val button = Button {
          action = Action("Click me") {
          	println("Button Clicked!")
          }
        }
        contents = button
      }
    }
\end{lstlisting}
As it is very common for buttons to have actions, it exists a convenient factory method in object Button.
This example is equivalent to the example above.
\begin{lstlisting}
val button = Button("Click me") {
	println("Button Clicked!")
}
\end{lstlisting}

\section{Scala Unit Testing}
\label{sec:scala_scalatest}

Scala Unit Testing with ScalaTest works in a similar way as traditional Unit testing with JUnit. 
But instead of working with annotations and \verb|assert| clauses, ScalaTest usually works with 
FlatSpec and Matchers. This allows for a more behaviour-driven development (BDD) style, where the 
test description can be read as part of the behaviour specification.

As a first step, one should create a trait or abstract class handling all mix-ins like does the following
trait UnitSpec:

\begin{lstlisting}
trait UnitSpec extends FlatSpec with Matchers with
  OptionValues with Inside with Inspectors with PrivateMethodTester
\end{lstlisting}

All test classes then extend this trait, which reduces the risk to forget a needed mix-in. 

The following test class shows the typical FlatSpec way of test definition which checks the result
using Matchers like \verb|shouldBe| or \verb|should not equal|:

\begin{lstlisting}
class ConnectionHandlerTest extends UnitSpec {
	val con = new ConnectionHandlerMock

	"A ConnectionHandler" should "get true if there is an internet connection" in {
		con.checkConnection shouldBe true
	}
	
	"A Socket" should "be returned" in {
		con.getSocket("localhost",8080) shouldBe a [Socket]
	}
	
	"A Proxy" should "be set" in {
		con.setProxy("localhost", 8080)
		con.getProxy shouldBe a [Proxy]
		it should not equal Proxy.NO_PROXY
	}
	[...]
\end{lstlisting}

%% file: anhang/A1-meetings.tex

\chapter{Meeting summary}
\label{chap:meeting}
\section*{Meeting 1 (20.02.2014)}
\label{sec:meeting_1}

The first meeting was a short kick-off to agree on some key ideas 
and to discuss important points to consider for planning and documentation:
\begin{itemize}
\item{Include article for book in planning}
\item{Mention all work in the report}
\item{Limit main report to 40-60 pages}
\item{Move explanation of Scala concepts to the appendix}
\item{Distinguish in working log who did which task}
\item{Remove academic title from cover page}
\item{Place report of project 2 in the appendix}
\end{itemize}

\section*{Meeting 2 (27.02.2014)}
\label{sec:meeting_2}

During the second meeting, various aspects of the report and also the defence were discussed.\\
Participants: Claude Fuhrer, Olivier Biberstein and Thomas Ender;\\ Patrick Vananti was absent due to military service.

\textbf{Expert} \\
The expert is presumably Dr. Joachim Wolfgang Kaltz, although it is now yet confirmed.

\textbf{Defence}
\begin{itemize}
\item{School-wide regulations apply}
\item{Spoken language can be German or French, also bilingual}
\item{Slides should be in English, especially with a bilingual presentation}
\item{The defence will be during the exam period (1 - 10 July 2014)}
\item{Details will be discussed during the meeting with the expert}
\end{itemize}

\textbf{Meetings with Olivier Biberstein}\\
Olivier Biberstein returns at least two times. A meeting is intended for each period.
\begin{itemize} 
\item{April: 5 - 7 April (weeks 14 - 15)}
\item{May: 7 - 10 May (week 19)}
\end{itemize}

\textbf{Report}\\
For the report, we shall use the package \verb|hyperref| to obtain click-able links. \\
In the bibliography, all web sources must contain the retrieval date.\\
Furthermore we discussed the extent and implementation of the API and its reference application:
\begin{itemize}
\item{Main goal is to create an environment favourable to easy extension}
\item{A common result format for all traces is a good idea. It may be simple Strings, but a more structured format like XML is better}
\item{Linking trace inputs with the obtained outputs allows to create a history view}
\item{Unit testing must be done with mock-up components, to be independent from network location}
\item{Performance profiling can be done using JVM tools like \verb|jconsole|}
\item{A trace to search for keywords should be added}
\item{Anonymization can be done using Vidalia}
\item{RegEx operations should be done using finite state automata, like with \verb|jflex| or \verb|antlr|}
\item{Olivier Biberstein will give information on useful tools}
\end{itemize}

\section*{Meeting 3 (13.03.2014)}
\label{sec:meeting_3}

During the third meeting, different aspects of the report were discussed.\\
Participants: Claude Fuhrer, Thomas Ender, and Patrick Vananti.\\

\textbf{Report}\\
For the report, we should put the references in the text and not at the end of a paragraph. We should describe the columns of the Table 2.1 and better explain the trace malicious activity. We should change the images so as to be more readable when printed and we should move glossary, list of figure and list of table at the beginning after the contents table.

\section*{Expert Meeting 1 (28.03.2014)}
\label{sec:expert_meeting_1}

The first expert meeting allowed us to introduce the project and the current state of work to the expert.\\
Participants: Dr. Joachim Wolfgang Kaltz, Claude Fuhrer, Thomas Ender, and Patrick Vananti.\\

\textbf{Next expert meeting} \\
The next meeting with the expert Dr. Kaltz is going to take place on the 30th April 2014 at 14:15

\textbf{Defence}\\
We fixed the date for the defence which will be held on 2nd July 2014 at 13:30.
Claude Fuhrer will take care of informing the school's authorities.

\textbf{Report and project}\\
We discussed the report and our work so far with the expert. Dr. Kaltz was mostly satisfied and we have the following tasks to fulfill:

\begin{itemize}
\item{Send the report of Project 2 to the expert}
\item{Prioritize and emphasize the goals of this project}
\item{Assess the state of work at every milestone and modify planning for the next part if necessary and}
\item{Note the analysis and decisions in the report}
\end{itemize}

\section*{Expert Meeting 2 (30.04.2014)}
\label{sec:expert_meeting_2}

The second meeting was used to discuss the current state of work to the expert.\\
Participants: Dr. Joachim Wolfgang Kaltz, Thomas Ender, and Patrick Vananti.\\

\textbf{Next expert meeting} \\
No further meeting planned

\textbf{Report and project}\\
We discussed the report and our work so far with the expert. Dr. Kaltz gave us his opinion on the report of project 2,
so we could pay attention during our current report. Overall he was satisfied and we have the following tasks to fulfill:

\begin{itemize}
\item{Check with Olivier Biberstein whether he will participate in the evalution}
\item{Explain the high-level intention for this project}
\item{Include the criteria for trace selection in the report}
\end{itemize}

Dr. Kaltz also offered to review a preliminary version of the report around the next milestone.

\section*{Meeting 4 (27.05.2014)}
\label{sec:meeting_4}

During the fourth meeting, different aspects of the report and the defence and other deliverables were discussed.\\
Participants: Claude Fuhrer, Thomas Ender\\

\textbf{Report}\\
For the tools used, the corresponding version used should be noted. This prevents difficulties if, in a future version, our configuration does not work.
Also a short usage help must be provided, noting which class has to be executed and with which parameters. 

\textbf{Poster and Book}\\
The poster for the "Finaltag" and the summary for the graduation thesis book have a mainly promotional nature. So we should focus on selling our work and
not write a too technical text. Better include some graphics to render it visually appealing.

\textbf{Defence}\\
To start off the presentation during the defence, it is beneficial to show a quick demonstration of the finished product, as to clarify what we are going 
to talk about. Following the formal presentation a longer and more detailed demonstration should follow. These demonstrations have to be prepared and tested
in advance, to avoid technical problems. Should there be a technical problem e.g. with the network connectivity, the results obtained during the preparation
can be used as a backup.

\textbf{Note:} Olivier Biberstein will attend the defence at Bienne.

%% file: anhang/A2-workinglog.tex

\chapter{Working log}

\label{chap:worklog}
\section*{Week 1 (8)}
\label{sec:worklog_01}
\begin{itemize}
\item{\textbf{VANANTI \& ENDER 17.02.2014 (4 hours) :} \\  Planning of Milestones and tasks }
\item{\textbf{VANANTI \& ENDER 18.02.2014 (6 hours) :} \\  Preparation of report, development environment, further planning, study Scala language}
\item{\textbf{VANANTI \& ENDER 19.02.2014 (6 hours) :} \\  Preparation of report, development environment, further planning, study Scala language}
\item{\textbf{VANANTI \& ENDER 20.02.2014 (6 hours) :} \\  Meeting with M. Biberstein and M. Fuhrer (Meeting 1).}
\end{itemize}
\section*{Week 2 (9)}
\label{sec:worklog_02}
\begin{itemize}
\item{\textbf{ENDER 25.02.2014 (8 hours) :} \\  Pre-selection of traces, preparation of report }
\item{\textbf{VANANTI 25.02.2014 (8 hours) :} \\  Military service (Self study Selection of traces)}
\item{\textbf{ENDER 26.02.2014 (7 hours) :} \\  Selection of traces}
\item{\textbf{VANANTI 26.02.2014 (7 hours) :} \\  Military service (Self study Scala Actors)}
\item{\textbf{ENDER 27.02.2014 (1 hour) :} \\  Meeting with M. Biberstein and M. Fuhrer (Meeting 2).}
\item{\textbf{ENDER 27.02.2014 (6 hours) :} \\  Learning Scala}
\item{\textbf{VANANTI 27.02.2014 (6 hours) :} \\  Military service (Self study Scala class and function)}
\end{itemize}
\section*{Week 3 (10)}
\label{sec:worklog_03}
\begin{itemize}
\item{\textbf{ENDER 04.03.2014 (8 hours) :} \\  Discussed package structure, detailed trace selection}
\item{\textbf{VANANTI 04.03.2014 (8 hours) :} \\  Discussed package structure, detailed trace selection}
\item{\textbf{ENDER 05.03.2014 (8 hours) :} \\  Rephrased API structure, experimented with workspace configuration}
\item{\textbf{VANANTI 05.03.2014 (8 hours) :} \\  Wrote and analysed a part of selected traces}
\item{\textbf{ENDER 06.03.2014 (8 hours) :} \\  Learning Scala, replacing scala.actors with akka.actor}
\item{\textbf{VANANTI 06.03.2014 (8 hours) :} \\  Learning Scala}
\end{itemize}
\section*{Week 4 (11)}
\label{sec:worklog_04}
\begin{itemize}
\item{\textbf{VANANTI \& ENDER 11.03.2014 (8 hours) :} \\  Discussion of API structure and general class structure, draft of trace explanation}
\item{\textbf{ENDER 12.03.2014 (8 hours) : }\\ Discussion of API/library distinction, revised trace sequence diagram, Scala (Akka) Actors explanation}
\item{\textbf{VANANTI 12.03.2014 (8 hours) :} \\  tbd}
\item{\textbf{ENDER 13.03.2014 (8 hours) :} \\  Meeting with M. Fuhrer, restructuring of API chapter}
\item{\textbf{VANANTI 13.03.2014 (8 hours) :} \\  Meeting with M. Fuhrer, restructuring of API chapter}
\end{itemize}
\section*{Week 5 (12)}
\label{sec:worklog_05}
\begin{itemize}
\item{\textbf{ENDER 18.03.2014 (8 hours) :} \\  First programming steps, created base structure with packages and traits}
\item{\textbf{VANANTI 18.03.2014 (8 hours) :} \\  Military service}
\item{\textbf{ENDER 19.03.2014 (8 hours) :} \\  Continued programming, added SD diagrams, studied network programming}
\item{\textbf{VANANTI 19.03.2014 (8 hours) :} \\  Military service}
\item{\textbf{ENDER 20.03.2014 (8 hours) :} \\  Introduced ConnectionHandler, added Proxy option, Fixed first expert meeting 
on Friday, 28th March 2014}
\item{\textbf{VANANTI 20.03.2014 (8 hours) :} \\  Military service}
\end{itemize}
\section*{Week 6 (13)}
\label{sec:worklog_06}
\begin{itemize}
\item{\textbf{VANANTI \& ENDER 25.03.2014 (8 hours) :} \\  Continued programming, fixed logical errors, code cleanup}
\item{\textbf{VANANTI \& ENDER 26.03.2014 (8 hours) :} \\  Continued programming, added test and mock frameworks, reworked report}
\item{\textbf{VANANTI \& ENDER 27.03.2014 (8 hours) :} \\  Continued programming}
\item{\textbf{VANANTI \& ENDER 28.03.2014 (1 hours) :} \\  Expert meeting 1}
\end{itemize}
\section*{Week 7 (14)}
\label{sec:worklog_07}
\begin{itemize}
\item{\textbf{ENDER 01.04.2014 (8 hours) :} \\  Continued test programming, added mock-up of ConnectionHandler, corrected spelling}
\item{\textbf{VANANTI 01.04.2014 (8 hours) :} \\  Wedding}
\item{\textbf{VANANTI \& ENDER 02.04.2014 (8 hours) :} \\  Programming of ServerInfo trace}
\item{\textbf{VANANTI \& ENDER 03.04.2014 (8 hours) :} \\  Programming of ServerInfo trace}
\end{itemize}
\section*{Week 8 (15)}
\label{sec:worklog_08}
\begin{itemize}
\item{\textbf{VANANTI \& ENDER 08.04.2014 (8 hours) :} \\  Rebuild of call hierarchy to use asynchronous method calls}
\item{\textbf{ENDER 09.04.2014 (8 hours) :} \\  Programming of SSL trace}
\item{\textbf{VANANTI 09.04.2014 (8 hours) :} \\  Modification of ServerInfo trace}
\item{\textbf{ENDER 10.04.2014 (8 hours) :} \\  Programming of SSL trace}
\item{\textbf{VANANTI 10.04.2014 (8 hours) :} \\  Programming of SSL trace}
\end{itemize}
\section*{Week Easter holiday (16)}
\label{sec:worklog_08a}
No work during holiday
\section*{Week 9 (17)}
\label{sec:worklog_09}
\begin{itemize}
\item{\textbf{VANANTI \& ENDER 22.04.2014 (8 hours) :} \\  Milestone "base version", finished programming of SSL and ServerInfo traces}
\item{\textbf{VANANTI \& ENDER 23.04.2014 (8 hours) :} \\  Updated Documentation, started programming of Whois and DNS traces}
\item{\textbf{VANANTI \& ENDER 24.04.2014 (8 hours) :} \\  Continued programming}
\end{itemize}
\section*{Week 10 (18)}
\label{sec:worklog_10}
\begin{itemize}
\item{\textbf{VANANTI \& ENDER 29.04.2014 (8 hours) :} \\  Continued programming, refactored package structure, added tests}
\item{\textbf{VANANTI \& ENDER 30.04.2014 (8 hours) :} \\  Expert meeting 2, finished Whois trace}
\item{\textbf{VANANTI \& ENDER 01.05.2014 (8 hours) :} \\  Updated documentation to reflect refactoring, started Metadata and Content traces}
\end{itemize}
\section*{Week 11 (19)}
\label{sec:worklog_11}
\begin{itemize}
\item{\textbf{ENDER 06.05.2014 (8 hours) :} \\  Continued programming of Metadata trace}
\item{\textbf{VANANTI 06.05.2014 (8 hours) :} \\  Continued programming of Content trace}
\item{\textbf{ENDER 07.05.2014 (8 hours) :} \\  Continued programming of Metadata trace, added tests}
\item{\textbf{VANANTI 07.05.2014 (8 hours) :} \\  Continued programming of Content trace, combine with Social Networks trace}
\item{\textbf{VANANTI \& ENDER 08.05.2014 (8 hours) :} \\  Milestone "Iteration 2", finished programming of Whois, DNS and Metadata traces}
\end{itemize}
\section*{Week 12 (20)}
\label{sec:worklog_12}
\begin{itemize}
\item{\textbf{ENDER 13.05.2014 (8 hours) :} \\  Started Malicious Activity trace, changed tests for Metadata trace }
\item{\textbf{VANANTI 13.05.2014 (8 hours) :} \\  Continued programming of Content and Social Networks trace}
\item{\textbf{ENDER 14.05.2014 (8 hours) :} \\  Continued programming of Malicious Activity trace, documentation update, planning update}
\item{\textbf{VANANTI 14.05.2014 (8 hours) :} \\  Continued programming of Content and Social Networks trace}
\item{\textbf{ENDER 15.05.2014 (8 hours) :} \\  Finished Malicious Activity Trace, tests}
\item{\textbf{VANANTI 15.05.2014 (8 hours) :} \\  Added combined Malicious Relations trace}
\end{itemize}
\section*{Week 13 (21)}
\label{sec:worklog_13}
\begin{itemize}
\item{\textbf{ENDER 20.05.2014 (8 hours) :} \\  Poster draft}
\item{\textbf{VANANTI 20.05.2014 (8 hours) :} \\  Continued programming of Malicious Relations trace}
\item{\textbf{ENDER 21.05.2014 (8 hours) :} \\  Programming of GUI structure}
\item{\textbf{VANANTI 21.05.2014 (8 hours) :} \\  Code review and optimization}
\item{\textbf{ENDER 22.05.2014 (8 hours) :} \\  Programming of GUI (main window)}
\item{\textbf{VANANTI 22.05.2014 (8 hours) :} \\  Programming of GUI (Result Panels)}
\end{itemize}
\section*{Week 14 (22)}
\label{sec:worklog_14}
\begin{itemize}
\item{\textbf{ENDER 26.05.2014 (4 hours) :} \\  Documentation update, Poster update}
\item{\textbf{ENDER 27.05.2014 (8 hours) :} \\  Meeting with M. Fuhrer, Programming of GUI (main window)}
\item{\textbf{VANANTI 27.05.2014 (8 hours) :} \\  Programming of GUI (Result Panels)}
\item{\textbf{ENDER 28.05.2014 (4 hours) :} \\  Programming of GUI (main window),Unit testing}
\item{\textbf{VANANTI 28.05.2014 (8 hours) :} \\  Programming of GUI (Result Panels)}
\item{\textbf{ENDER 29.05.2014 (8 hours) :} \\  Unit testing, Poster update}
\item{\textbf{VANANTI 29.05.2014 (8 hours) :} \\  Programming of GUI (Result Panels)}
\end{itemize}
\section*{Week 15 (23)}
\label{sec:worklog_15}
\begin{itemize}
\item{\textbf{VANANTI \& ENDER 03.06.2014 (8 hours) :} \\  Documentation update}
\item{\textbf{VANANTI \& ENDER 04.06.2014 (8 hours) :} \\  Documentation update}
\item{\textbf{VANANTI \& ENDER 05.06.2014 (8 hours) :} \\  Meeting with M. Fuhrer, Documentation update}
\end{itemize}
\section*{Week 16 (24)}
\label{sec:worklog_16}
\begin{itemize}
\item{\textbf{VANANTI \& ENDER 11.06.2014 (8 hours) :} \\  Proof reading, creating Finaltag presentation}
\item{\textbf{VANANTI \& ENDER 12.06.2014 (8 hours) :} \\  Printing and preparing the deliverables}
\item{\textbf{VANANTI \& ENDER 13.06.2014 (8 hours) :} \\  Finaltag}
\end{itemize}

%% file: A4-projecttwo.tex

\chapter{Report of Project 2}
\label{chap:projecttwo}

\includepdf[pages=-,scale=1]{./anhang/A4-projecttwo.pdf} 